\documentclass[lettersize,journal]{IEEEtran}
\usepackage{amsmath,amssymb,amsfonts}
\usepackage[utf8]{inputenc}
\usepackage{algorithmic}
\usepackage{algorithm}
\usepackage{array}
\usepackage[caption=false,font=scriptsize,labelfont=sf,textfont=sf]{subfig}
\usepackage{textcomp}
\usepackage{stfloats}
\usepackage{url}
\usepackage{verbatim}
\usepackage{graphicx}
\usepackage{cite}
\usepackage{amssymb}
\usepackage{setspace}
\usepackage{lineno} 
\usepackage{cases}
\usepackage{float}  
\usepackage{titlesec}
\usepackage{textcomp}
\usepackage{bm,bbold}
\usepackage{xcolor}
\usepackage{siunitx}
\newcommand\bw{\ensuremath{{\bold w}}}

\newcommand\bW{\ensuremath{{\bold W}}}
\newcommand\bJ{\ensuremath{{\bold J}}}

\newcommand\bx{\ensuremath{{\bm x}}}

\newcommand\bG{\ensuremath{{\bold G}}}
\newcommand\bh{\ensuremath{{\bm h}}}
\newcommand\bH{\ensuremath{{\bold H}}}

\newcommand\be{\ensuremath{{\bm e}}}
\newcommand\bz{\ensuremath{{\bm z}}}

\newcommand\bp{\ensuremath{{\bm p}}}

\newcommand\bC{\ensuremath{{\bold C}}}
\newcommand\bc{\ensuremath{{\bm c}}}
\newcommand\ba{\ensuremath{{\bm a}}}

\newcommand\bA{\ensuremath{{\bold A}}}

\newcommand\bb{\ensuremath{{\bm b}}}
\newcommand\bg{\ensuremath{{\bm g}}}
\newcommand\bB{\ensuremath{{\bold B}}}

\newcommand\bF{\ensuremath{{\bold F}}}

\newcommand\bd{\ensuremath{{\bm d}}}
\newcommand\bD{\ensuremath{{\bold D}}}

\newcommand\bu{\ensuremath{{\bm u}}}
\newcommand\bv{\ensuremath{{\bm v}}}

\newcommand\bPhi{\ensuremath{{\bm \Phi}}}

\newcommand\bs{\ensuremath{{\bm s}}}

\newcommand\bE{\ensuremath{{\bold E}}}

\newcommand{\diag}{\mathrm{diag}}

\newcommand{\bI}{{\bold I}}

\begin{document}

\title{ Two-Stage Coded-Sliding Beam Training and QoS-Constrained Sum-Rate Maximization for SIM-Assisted Wireless Communications
}

\author{Qian Zhang,~\IEEEmembership{Graduate Student Member,~IEEE,}
		Ju Liu,~\IEEEmembership{Senior Member,~IEEE,}
		Yao Ge,~\IEEEmembership{Member,~IEEE,}
		Yufei Zhao,
		Wali Ullah Khan,~\IEEEmembership{Member,~IEEE,}
		Zheng Dong,~\IEEEmembership{Member,~IEEE,}
		Yong Liang Guan,~\IEEEmembership{Senior Member,~IEEE,}
		and 
		Chau Yuen,~\IEEEmembership{Fellow,~IEEE}
\thanks{
	This work was supported in part by the National Natural Science Foundation of China under Grant 62071275;
	in part by the China Scholarship Council;
	in part by MOE (Ministry of Education, Singapore), under MOE Tier 2 Award number T2EP50124-0032;
	in part by A*STAR under the RIE2025 Industry Alignment Fund–Industry Collaboration Projects (IAF-ICP) Funding Initiative (Award: I2501E0045), as well as cash and in-kind contribution from the industry partner(s).
	The corresponding authors: Ju Liu; Zheng Dong. }
\thanks{Qian Zhang is with School of Information Science and Engineering, Shandong University, Qingdao 266237, China, and with School of Electrical and Electronic Engineering, Nanyang Technological University, Singapore (e-mail: qianzhang2021@mail.sdu.edu.cn).}
\thanks{Ju Liu and Zheng Dong are with School of Information Science and Engineering, Shandong University, Qingdao 266237, China (e-mail: juliu@sdu.edu.cn; zhengdong@sdu.edu.cn).}
\thanks{Wali Ullah Khan is with the Interdisciplinary Centre for Security, Reliability, and Trust (SnT), University of Luxembourg, 1855 Luxembourg City,	Luxembourg (e-mail: waliullah.khan@uni.lu).}
\thanks{Yao Ge is with the AUMOVIO-NTU Corporate Lab, Nanyang Technological University, Singapore 639798 (e-mail: yao.ge@ntu.edu.sg).}
\thanks{Yufei Zhao, Yong Liang Guan, and Chau Yuen are with School of Electrical and Electronic Engineering, Nanyang Technological University, Singapore 639798 (e-mail: yufei.zhao@ntu.edu.sg; EYLGuan@ntu.edu.sg; chau.yuen@ntu.edu.sg).}
}

\markboth{}%
{Shell \MakeLowercase{\textit{et al.}}: A Sample Article Using IEEEtran.cls for IEEE Journals}


\maketitle

\begin{abstract}
Stacked intelligent metasurfaces (SIM) provide a cost-effective and scalable solution for large-scale antenna communications. 
However, efficient channel state information acquisition and phase shift optimization remain critical challenges. 
In this paper, we develop a unified framework of low-complexity algorithms for SIM-assisted communication systems to address these issues.
Specifically, we propose a generalized two-step codebook construction (TSCC) method that leverages two-dimensional angular-domain decoupling to transform planar array beamformer design into two independent one-dimensional linear array beamformer design problems, efficiently solved via the Gerchberg–Saxton algorithm and our proposed majorization–minimization-based proximal-distance (PDMM) algorithm. 
We further develop a two-stage coded-sliding beam training (TSCSBT) method for low-overhead and high-accuracy beam training, where error-correcting codes are embedded in the first-stage training to enhance robustness against noise, and sliding sampling is subsequently performed around the matched angular samples to improve angular resolution. The proposed framework is further extended to multi-path user channels.
Finally, a variable decoupling-based block successive upper bound minimization (VD-BSUM) algorithm is proposed to directly solve the QoS-constrained sum-rate maximization problem through closed-form iterative updates with substantially reduced computational complexity.
Simulation results demonstrate the effectiveness of the proposed methods in achieving precise beam pattern realization, improved beam training accuracy and angular resolution, and enhanced sum-rate performance.

\end{abstract}

\vspace{-0.2cm}

\vspace{-0.6cm}
\section{Introduction}\vspace{-0.15cm}
The rapid growth of wireless communications has given rise to the steadily increasing the requirement for higher data rates, lower latency, and massive connectivity~\cite{Jiang20246G}. 
To satisfy these unprecedented requirements, researchers have explored novel transceiver architectures and low-overhead techniques to enhance the effectiveness and reliability of communications.
In the past several decades, multiple-input multiple-output (MIMO) technology has emerged as a revolutionary technological innovation by equipping multiple antennas at the transceiver to exploit multiplexing gain and spatial diversity to enhance spectral efficiency (SE) and link reliability~\cite{Bjornson2017MssiveMIMO}.
With further technological evolution, wireless communications are moving towards higher frequencies and larger-scale MIMO.
However, the full-digital beamformer-based MIMO technology makes it difficult to realize large-scale array evolution due to the expensive radio frequency (RF) chains.
To reduce the hardware overhead, a hybrid MIMO architecture has been proposed, which combines a portion of the RF chain with low-cost analog phase shifters to achieve a digital-analog hybrid beamformer architecture~\cite{Ahmed2018Hybrid_MIMO}.
Unfortunately, due to the restricted RF chains and complicated structural design, hybrid MIMO still has some limitations in enhancing the SE.

Stacked intelligent metasurfaces (SIM) have been recently proposed for the purpose of achieving low-cost and high-gain communications~\cite{An2024SIM,An2023SIM_JSAC,Niu2025TWO_SIM,Liu2025SIM_WCM,Wang2024Globecom}.
SIM, consisting of multiple programmable transmissive metasurfaces with a large number of low-cost meta-atoms, has its prototype and variants detailed in~\cite{Liu2025SIM_WCM,Wang2024Globecom}. 
By stacking multiple layers, SIM enables multidimensional wave manipulation through inter-layer electromagnetic interactions, achieving greater spatial control than conventional single-layer RIS~\cite{Guo2020RIS,Huang2019RIS,Zhou2020RIS,Zhang2023RIS}. Compared with conventional MIMO architectures, SIM provides a low-cost solution with flexible beam control, energy focusing, and enhanced effective rank~\cite{An2024SIM,An2023SIM_JSAC}, making it a promising candidate for systems requiring efficient spatial multiplexing. 
The multi-layer design further improves beamforming versatility and multiplexing gain, but introduces additional phase-shift design complexity due to inter-layer coupling.
The phase shifts of each SIM layer can be properly configured to achieve flexible modulation of electromagnetic waves and efficient beamforming tasks in the wave domain to enhance SE, energy efficiency, and link reliability~\cite{Shi2025SIM_EE,Niu2025TWO_SIM,An2025SIM_downlink}.
Currently, SIM has been demonstrated to achieve high gain in many scenarios, such as secure transmission communications~\cite{Niu2025TIFS}, direction-of-arrival estimation~\cite{An2024SIM_DoA}, and ISAC systems~\cite{Zhang2025SIM_ISAC}. 
Therefore, SIM technology is a promising enabler for the future evolution of high-frequency communication and large-scale MIMO systems. 
Nonetheless, large-scale SIM systems incur a substantial increase in the pilot overhead for channel estimation, making the acquisition of channel state information (CSI) challenging ~\cite{Yao2024SIM_CE,Han2021LSMIMO,Zheng2025CBT}. 
Therefore, efficient CSI acquisition is crucial in large-scale SIM systems.

To this end, beam training is a practical way to obtain user CSI~\cite{Xiao2016HCodebook_Criteria,Chen2020Codeword,Zheng2025CBT}. 
Beam training is typically performed using a predefined codebook composed of candidate beamforming vectors. 
The transmitter sequentially transmits the pilot signal precoded using codewords and then obtains the CSI by feedback from users.
Generally, beam training is categorized into two types, i.e., exhaustive beam training (EBT) and hierarchical beam training (HBT).
EBT enables high-gain narrow-beam sweeping over the entire angular domain and therefore tends to obtain accurate CSI~\cite{Alkhateeb2015EBT,Sun2019EBT}.
However, the training overhead of EBT in large-scale antenna systems is generally unaffordable.
Binary HBT employs a layer-by-layer approach to narrow down the possible directions of the target user, where each layer performs two wide-beam scans~\cite{Xiao2016HCodebook_Criteria,Qi2020HCodebook,Zheng2025CBT}. 
As a result, HBT significantly reduces the training overhead and is a suitable way to obtain CSI in large-scale antenna systems.
Unfortunately, if HBT misidentifies the user's direction at any layer, the resulting error may propagate and lead to inaccurate CSI acquisition~\cite{Zheng2025CBT,Chen2025CBT_RIS}.
To enhance training accuracy, Zheng {\it et al.} introduced a concept of coded beam training (CBT) in~\cite{Zheng2025CBT}, which embeds error-correcting codes into the HBT framework, enabling self-correction in the training process. 
Chen {\it et al.} further extended the CBT framework to reconfigurable intelligent surface (RIS)-assisted wireless systems~\cite{Chen2025CBT_RIS}.
Although beam training has been widely studied, codebook design and training schemes for SIM-assisted systems remain largely unexplored. 
Moreover, the simplification of HBT in the two-dimensional (2-D) angular domain has received limited attention.

On the other hand, beamforming vectors can be efficiently designed based on CSI to enhance multiuser SE. 
In this case, beamforming optimization through multiuser sum-rate maximization (SRM) is a common way and has also received much attention~\cite{Bjornson2017MssiveMIMO,An2025SIM_downlink,Papazafeiropoulos2024SIM_SCSI,Liu2025SIM_DRL}.
Currently, the considered SRM problems are mainly categorized into two cases: 1) disregarding the quality-of-service (QoS) constraints for each user, and 2) guaranteeing the QoS for each user.
In general, the SRM problem without QoS constraints can be solved in a simple and efficient way, but it is prone to serious skewed resource allocation at low signal-to-noise ratio (SNR), i.e., all the resources are fully allocated to a single user with the best channel quality, resulting in the blocking of other users with poor channel quality. 
Therefore, the communication quality for each user can be guaranteed by adding QoS constraints, but this will lead to high complexity and difficulty in solving the problem.
The solution of the current SRM problem without QoS constraints has been studied in SIM-assisted multiuser systems~\cite{An2025SIM_downlink,Papazafeiropoulos2024SIM_SCSI,Liu2025SIM_DRL}.
However, to the best of our knowledge, the SRM problem with QoS constraints has not been studied in SIM-assisted communication systems.
In addition, the commonly used approach to solve the QoS-constrained SRM problem is to convert it into a convex form using the weighted minimum-mean-square-error (WMMSE) method~\cite{shi2011wmmse}, fractional programming (FP) algorithm~\cite{shen2018FP}, successive convex approximation (SCA) algorithm~\cite{Razaviyayn2014SCA} or the semi-definite relaxation (SDR) algorithm~\cite{luo2010sdr} and then solve it using CVX~\cite{grant2014cvx}.
These approaches typically results in high computational complexity.
Especially with multiple layers and multiple meta-atoms, the computational complexity generated by these methods in large-scale SIM systems is arguably incalculable.
Therefore, the development of a low-complexity algorithm to solve the multiuser SRM problem with QoS constraints is urgently needed.

To address the above challenges, we propose a beam training method based on low-complexity codebook construction and an efficient beamformer optimization algorithm in SIM-assisted communication systems. Specifically, our main contributions are summarized as follows. \vspace{-0.05cm}
\begin{itemize}
	\item[$\bullet$] We design a generalized two-step codebook construction (TSCC) method for low-complexity SIM codebook design, where a novel 2-D angular-domain decoupling approach converts the planar array beamformer design into two independent linear array beamformer designs, which are efficiently solved via the Gerchberg–Saxton (GS) algorithm. Furthermore, a majorization–minimization-based proximal distance (PDMM) algorithm is developed to generate practical SIM codewords with closed-form updates and guaranteed convergence to the Karush–Kuhn–Tucker (KKT) points.
	
	\item[$\bullet$] We present a novel two-stage coded-sliding beam training (TSCSBT) method that combines 2-D angular-domain decoupling-based binary HBT with error-correcting codes for enhanced noise robustness, and sliding sampling for improved angular resolution. Furthermore, we extend the framework to a multi-path two-stage beam training (MP-TSBT) scheme, enabling efficient beam training for multi-path users, which has not been previously explored.

	\item[$\bullet$] We propose a novel optimization framework, termed the variable-decoupling-based block successive upper bound minimization (VD-BSUM) algorithm, for QoS-constrained SRM problems. In the outer layer, BSUM-based variable decoupling transforms the original problem into tractable subproblems. In the inner layer, a low-complexity PDMM algorithm independently handles each QoS constraint by converting non-convex quadratic constraints into single-variable affine forms with increasing penalty dual decomposition (IPDD)-based closed-form projections, enabling efficient updates in closed form.

	\item[$\bullet$]Our simulation results demonstrate the effectiveness of the proposed methods. The proposed TSCC algorithm accurately realizes the desired beam pattern. The proposed TSCSBT method achieves higher training accuracy than independent binary HBT along azimuth and elevation, with improved angular resolution and user rate compared to conventional EBT. The extended MP-TSBT accurately identifies multiple propagation paths. Finally, the VD-BSUM algorithm achieves high performance with substantially lower computational complexity compared to SCA and SDR algorithms.

\end{itemize}

{\it{Notation:}} 
$||\cdot||_2$ and $|\cdot|$ express the Euclidean norm and absolute value, respectively.
$\bx \sim {\cal{CN}}\left( \bm 0, \bm\Gamma \right)$ represents that $\bx$ follows the circularly symmetric complex Gaussian (CSCG) distribution with mean vector $\bm 0$ and covariance matrix $\bm\Gamma$.
$(\cdot)^{*}$, $(\cdot)^{\rm T}$, $(\cdot)^{\rm H}$, and $(\cdot)^{-1}$ denote the conjugate, transpose, and Hermitian transpose, matrix inverse, respectively.
$\mathbb{C}^{M \times N}$ denotes $M \times N$ complex matrix space.
$\odot$ and $\otimes$ are the Hadamard product and the Kronecker product, respectively.
$\bold{I}_K$ is a $K \times K$ identity matrix.
${\rm diag}(\cdot)$ is the diagonalization operation.
$\Re\{\cdot\}$ denotes the real part.
$\cap$ is the intersection operation.
${\rm mod}(\cdot,\cdot)$ denotes a remainder operation.
$[\ba]_n$ denotes the $n$-th entry of the vector $\ba$.

\begin{figure*}[t]	
	\centering \includegraphics[width=\linewidth]{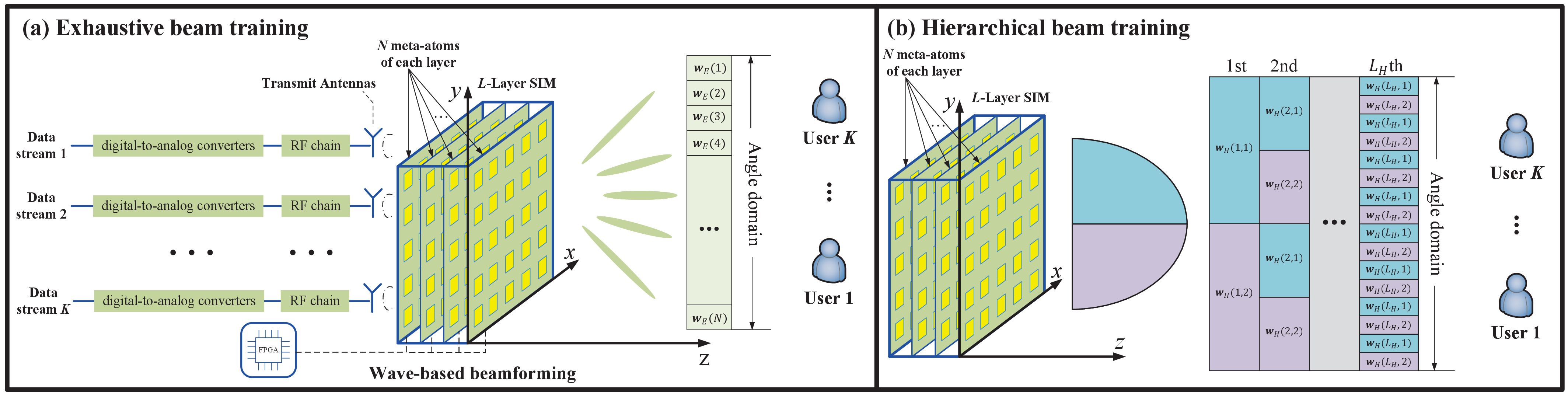}
	\vspace{-0.6cm}
	\caption{Traditional beam training frameworks. (a) Exhaustive beam training; (b) Hierarchical beam training.}
	\vspace{-0.55cm}
	\label{fig:System_Model_BeamTraining}
\end{figure*}

\vspace{-0.4cm}
\section{System Model and Background} \vspace{-0.1cm}
In this section, we first introduce the signal and channel models for the SIM-assisted communication system and formulate the QoS-constrained SRM problem, noting that phase-shifter optimization for SIM under this problem has not yet been studied. Finally, we present the basic principles of conventional EBT and HBT for acquiring user CSI.

\vspace{-0.4cm}
\subsection{System Model} \vspace{-0.05cm}
We consider a SIM-assisted downlink communication system in which the $M$-antenna base station (BS) transmits $K$ data streams, with the assistance of the $L$-layer SIM consisting of $N$ elements per layer, serves $K$ single-antenna users.
Without loss of generality, we assume that $M = K$ and that each antenna performs the modulation and detection of a single stream~\cite{An2024SIM,An2023SIM_JSAC,An2025SIM_downlink}.
In such a system, the BT method is used to obtain the CSI of each user, and then we optimize the wave-based beamforming vector to achieve high SE multiuser communication.
Specifically, the transmitted signal at the BS is $\bx = \sum_{k=1}^{K} \bG \bold{w}_k^1 \sqrt{\overline p_k} s_k $, where $\bold{w}_k^1 \in \mathbb{C}^{N\times 1}$ denotes the channel between the $k$-th BS antenna and first layer of SIM.
As a result, the received signal at user $k$ is given by \vspace{-0.1cm}
\begin{equation}
	\begin{split}
		y_k = \bh_k^{\rm H} \bx + n_k,~k\in {\cal K} \triangleq \left\{ 1,2,\dots,K \right\},
	\end{split}
\end{equation}
\vspace{-0.5cm}

\noindent where $\overline p_i\geq 0$ is the allocated power for user $i$ such that $\sum_{i=1}^{K} \overline p_i \leq P_{\rm max}$, $P_{\rm max}$ is the transmit power budget, $\bh_k \in \mathbb{C}^{N\times 1}$ denotes the channel from the $L$-th layer of SIM to user $k$, $n_k \sim {\cal CN}\left( 0, \sigma_k^2 \right)$ denotes the noise at user $k$, $\bs = \left[ s_1,s_2,\dots,s_K \right]^{\rm T}$ denotes the data symbol for users such that $\mathbb{E}\left[ \bs \bs^{\rm H} \right] = \bI_K$,
and $\bG$ is the wave-based beamforming matrix achieved by SIM as follows
\begin{equation}
	\begin{split}
		\bG = \bm\Phi^L \bW^L \bm\Phi^{L-1} \cdots \bm\Phi^2 \bW^2 \bm\Phi^1,
	\end{split}
\end{equation}
\vspace{-0.5cm}

\noindent where $\bPhi^l \in \mathbb{C}^{N\times N}$ is the diagonal phase shift matrix of the $l$-th layer of SIM and $\bW^l \in \mathbb{C}^{N\times N}$ is the channel matrix between $(l-1)$-th layer and $l$-th layer of SIM.

Based on the Rayleigh-Sommerfeld diffraction theory~\cite{Lin2018Diffractive}, the $(n,n^{'})$-th entry $w_{n,n^{'}}^l$ of $\bW^l$ is given by \vspace{-0.2cm}
\begin{equation}\label{layer_channel}
	\begin{split}
		w_{n,n^{'}}^l = \frac{d_xd_y \cos\varphi_{n,n^{'}}^l \left( \lambda - j 2\pi d_{n,n^{'}}^l \right) }{2\pi \left(d_{n,n^{'}}^l\right)^2 \lambda} e^{j 2\pi d_{n,n^{'}}^l / \lambda },
	\end{split}
\end{equation}
\vspace{-0.35cm}

\noindent for any $n\in {\cal N}$ and $l\in {\cal L} \setminus \{1\}$, where ${\cal L} \triangleq \left\{ 1,2,\dots,L \right\} $, ${\cal N} \triangleq \left\{ 1,2,\dots,N \right\} $, $\lambda$ denotes the wavelength, $d_{n,n^{'}}^l$ denotes the transmission distance, $\varphi_{n,n^{'}}^l$ is the angle between the signal propagation direction and the normal direction at the $(l-1)$-th layer, and $d_x\times d_y$ is the size of each meta-atom.
$\bold w_k^1$ can also be obtained by~\eqref{layer_channel}, which is given by \vspace{-0.2cm}
\begin{equation}
	\begin{split}
		[\bold w_k^1]_n = \frac{d_xd_y \cos\varphi_{k,n}^1 \left( \lambda - j 2\pi d_{k,n}^1 \right) }{2\pi \left(d_{k,n}^1\right)^2 \lambda} e^{j 2\pi d_{k,n}^1 / \lambda },
	\end{split}
\end{equation}
\vspace{-0.4cm}

\noindent where $\varphi_{k,n}^1$ is the angle between the signal propagation direction (from $k$-th antenna to $n$-th meta-atom of the first layer of SIM) and the normal direction at the transmit antenna array, $d_{k,n}^1$ denotes the distance between $k$-th antenna and $n$-th meta-atom of the first SIM layer.

Furthermore, due to the pronounced scattering attenuation at high frequencies (e.g., millimeter-wave and terahertz bands), the gain of the line-of-sight (LoS) path becomes substantially higher than that of the non-line-of-sight (NLoS) paths. Consequently, the LoS component dominates high-frequency signal propagation and serves as the primary contributor to data transmission~\cite{Chen2025CBT_RIS,Zheng2025CBT,Zhang2025XL_RIS}.
Therefore, we consider aligning the signal energy to the LoS path between the SIM and users during beam training. 
Based on this, we denote the channel $\bh_k$ as follows \vspace{-0.25cm}
\begin{equation}
	\begin{split}
		\bh_k = \alpha_k \ba(\theta_k, \varphi_k),
	\end{split}
\end{equation} 
\vspace{-0.55cm}

\noindent where $\alpha_k$ denotes the large-scale propagation loss with respect to user $k$, $\theta_k$ and $\varphi_k$ are the elevation angle and the azimuth angle of the $k$-th user, and 
\begin{equation*}
	\begin{split}
		\ba(\theta_k, \varphi_k) = 
		&[ 1, e^{-j\pi {\rm sin}(\theta_k){\rm sin}(\varphi_k)} , \dots,e^{-j\pi (N_1 - 1) {\rm sin}(\theta_k) {\rm sin}(\varphi_k)}
		]^{\rm T} \\
		& \otimes [ 1, e^{-j\pi {\rm cos}(\theta_k)}, \dots,e^{-j\pi (N_2 - 1)  {\rm cos}(\theta_k)}
		]^{\rm T} \big/ \sqrt{N},
	\end{split}
\end{equation*}
where $N = N_1 \times N_2$ in which $N_1$ and $N_2$ denote the number of meta-atoms along the $x$-axis and $y$-axis, respectively.

Using the proposed codebook design method, the resulting codebook facilitates the beam training procedure to obtain the corresponding channel information, which is described in Sections III and IV. After acquiring the users’ CSI, we aim to optimize the SIM phase shifters to maximize the multiuser sum rate under QoS constraints, which can be mathematically expressed as  \vspace{-0.2cm}
\begin{equation} \label{WSRM_QoS}
	\begin{split}
		&\max_{ \{ \bm\Phi^l \}_{l=1}^L, \overline\bp } ~~ \sum_{k=1}^{K} R_k  \\
		&\mbox{s.t.}~	
		{\cal C}^G: \bG = \bm\Phi^L \bW^L \bm\Phi^{L-1} \cdots \bm\Phi^2 \bW^2 \bm\Phi^1, \\
		&~ {\cal C}^\phi: \bPhi^l = \diag(\bm\phi^l),~\left|\phi_n^l\right| = 1,~l\in {\cal L},\,n \in {\cal N}, \\
		&~ {\cal C}^p: \sum_{k=1}^{K} \overline p_k \leq P_{\rm max},~\overline p_k\geq0, k \in {\cal K}, \\
		&~ {\cal C}_{k}^R: \frac{ | \bh_k^{\rm H} \bG \bw_k^1  |^2 \overline p_k }{\sum_{i=1,i\neq k}^{K}| \bh_k^{\rm H} \bG \bw_i^1  |^2  \overline p_i + \sigma_k^2 } \geq \gamma_{\rm th}, k \in {\cal K},
	\end{split}
\end{equation}
\vspace{-0.3cm}

\noindent where $\overline\bp = [\overline p_1,\dots,\overline p_K]^{\rm T}$, $\gamma_{\rm th} = 2^{R_{\rm th}} - 1$ and $R_{\rm th}$ denote the user rate threshold, and $R_k$ is the achievable rate of $k$-th user as follows \vspace{-0.15cm}
\begin{equation}
	\begin{split}
		R_k = {\rm log}_2\,\left(1 + \frac{ | \bh_k^{\rm H} \bG \bw_k^1  |^2 \overline p_k }{\sum_{i=1,i\neq k}^{K}| \bh_k^{\rm H} \bG \bw_i^1  |^2  \overline p_i + \sigma_k^2 } \right).
	\end{split}
\end{equation}

\vspace{-0.15cm}
\noindent Constraints ${\cal C}^G$ and ${\cal C}^\phi$ ensure that the beamforming matrix adheres to the constant modulus phase-shift structure required by SIM. Constraint ${\cal C}^p$ guarantees non-negative power allocation for each data stream and ensures that the total transmit power does not exceed the budget $P_{\rm max}$. 
Constraint ${\cal C}^{k}$ enforces a minimum achievable rate $R_{\rm th}$ for each user, preventing service outage for users with poor channel conditions, especially under low transmit power conditions.


\vspace{-0.45cm}
\subsection{Traditional Beam Training Frameworks} \vspace{-0.1cm}
Beam training is generally considered as an efficient way to acquire CSI.
At present, there are two commonly used beam training methods, i.e., exhaustive beam training (EBT) and binary HBT, as shown in Fig.~\ref{fig:System_Model_BeamTraining}.
For simplicity, we refer to the elevation and azimuth directions as the  $\theta$ and $\varphi$ directions, respectively, in the following beam training description.

\subsubsection{Exhaustive Beam Training}
As shown in Fig.~\ref{fig:System_Model_BeamTraining}$\,$(a), the BS loads codewords in exhaustive codebooks to transmit directional narrow beams and sequentially searches for the angles corresponding to all codewords.
The codebook for beam training over the entire angular domain is denoted as $\left\{\bm w_E(1),\bm w_E(2),\dots, \bm w_E(N)  \right\}$, where $\bm w_E(i)$ represents the $i$-th codeword corresponding to a specific $\theta$ and $\varphi$ direction.
Therefore, in planar array-based BS systems, traditional EBT employs a 2-D grid search strategy, resulting in high training overhead.
In our considered systems, the number of codewords is determined by the SIM-tunable spatial angular resolution and coverage range, which generally leads to an unacceptable overhead.
Especially when the SIM size becomes larger, the EBT method will be difficult to realize in practical systems.

\begin{figure}[t]	
	\centering \includegraphics[width=0.68\linewidth]{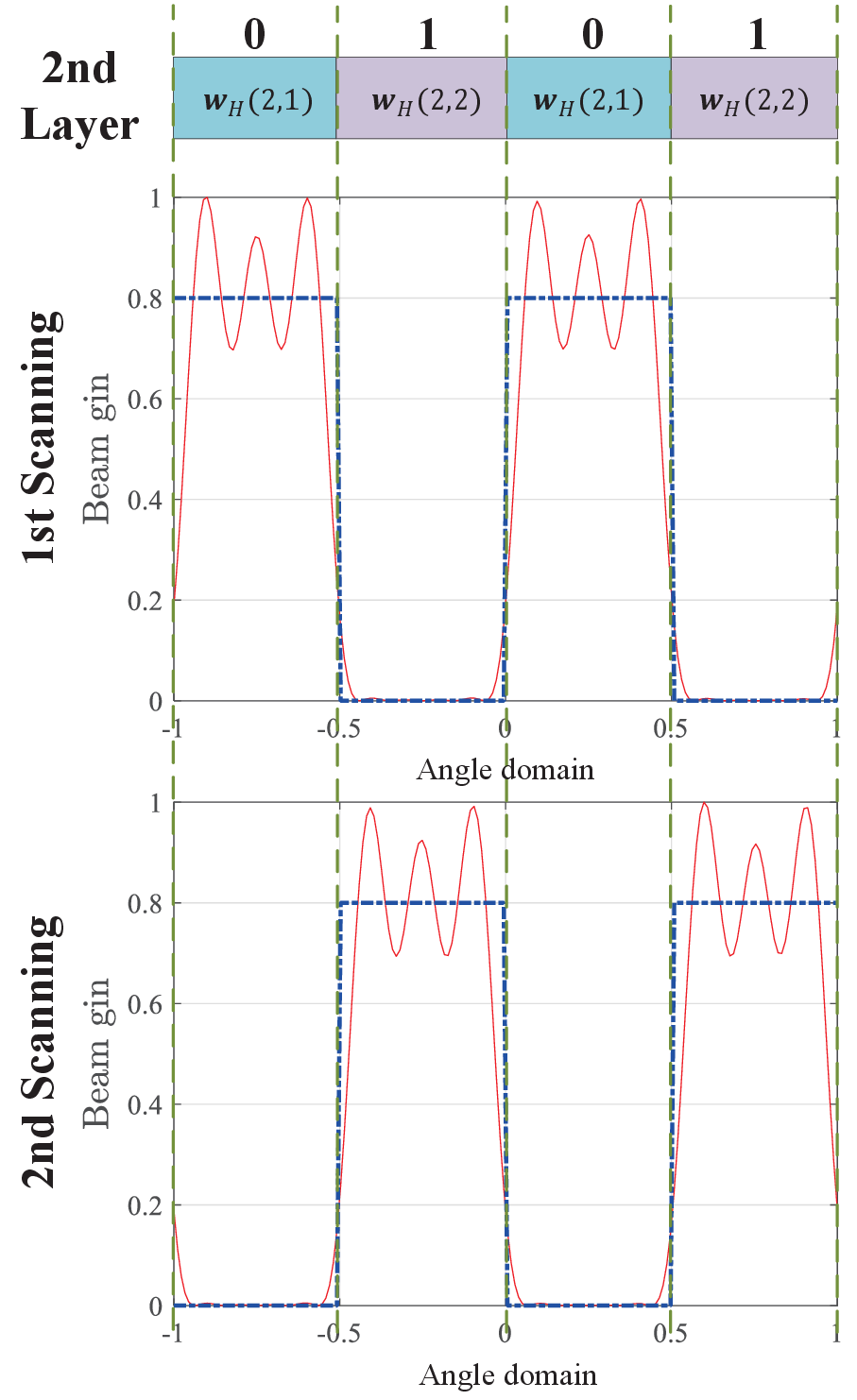}
	\vspace{-0.35cm}
	\caption{Example of the beam pattern corresponding to the second layer of codewords in the hierarchical codebook.}
	\vspace{-0.5cm}
	\label{fig:Codebook_waveform}
\end{figure}
\subsubsection{Hierarchical Beam Training}
Binary HBT scheme can effectively decrease training overhead compared to EBT~\cite{Shi2020PDD,Chen2025CBT_RIS,Zheng2025CBT}. 
As shown in Fig.~\ref{fig:System_Model_BeamTraining}$\,$(b), we apply HBT based on binary search to obtain the user CSI.
Specifically, in the first layer, the entire angular domain is divided into two regions for beam scanning. 
In each subsequent layer, every sub-region from the previous layer is further divided into two smaller regions, progressively refining the angular resolution.

Two beam scans are used for each layer, i.e., each layer corresponds to two codewords.
We define the codewords corresponding to the first and second beam scans of the $i$-th layer as $\bm w_H(i,1)$ and $\bm w_H(i,2)$, respectively.
In order to clearly elaborate the HBT, we take the second layer of BT corresponding to codeword $\left\{ \bm w_H(2,1), \bm w_H(2,2) \right\}$ as an example, as shown in Fig.~\ref{fig:Codebook_waveform}.
When the power received at the user in the first beam scanning is greater than the power received in the second beam scanning, then the feedback symbol is 0, and vice versa the feedback symbol is 1.
Suppose that after HBT based on a 4-layer codebook, we obtain the feedback symbol $\bm u$ is $\left[ 0,1,0,1 \right]$.
Then, we can finally obtain the user angle range markers as $2^3\times 0 + 2^2 \times 1 + 2^1 \times 0 + 2^0\times 1 + 1 = 6$, which corresponds to $\bw_E(6)$ in the exhaustive codebook.

{\it Remark 1:
Although the HBT method effectively reduces overhead, it is prone to error-prone feedback under low SNR, especially during multi-layer searches, which can lead to inaccurate CSI acquisition.
For instance, the intended feedback symbol is $\left[ 0,1,0,1 \right]$, but due to noise, it may be received as $\left[ 0,1,0,0 \right]$, causing the selected codeword to deviate from the correct $\bw_E(6)$ to an incorrect $\bw_E(5)$.
Therefore, enhancing the noise robustness of binary HBT remains crucial.
}

\vspace{-0.45cm}
\section{Two-Step Codebook Construction Method} \vspace{-0.1cm}
In this section, we propose a 2-D angle-domain decoupling method to enable the design of low-complexity codebook construction for beam training.
Specifically, we transform the design of a desired planar array beamformer into the Kronecker product of two linear array beamformers based on angular-domain decoupling.
By approximating the resulting desired beamformer with a wave-based SIM beamformer, the practical SIM codebook is realized as the set of SIM phase shifters employed for beam training.

\vspace{-0.5cm}
\subsection{TSCC Method and Angle-Domain Decoupling} \vspace{-0.1cm}
An effective codebook design is essential for ensuring successful beam training and achieving high training accuracy. 
Consequently, in SIM-assisted communication systems, the development of a well-structured and efficient codebook construction method is of paramount importance.
However, existing research has yet to explore the design of codebooks specifically for SIM-assisted communication systems.
Therefore, we aim to develop an efficient codebook construction method tailored for SIM systems.
A widely adopted approach for codebook construction is to approximate the desired beam pattern using beam patterns generated through angle-domain sampling, thereby constructing the desired codebook. 
Following this principle, the codebook design problem in SIM systems can be formulated as follows \vspace{-0.1cm}
\begin{equation} \label{coded_codebook}
	\begin{split}
		\min_{ \{ \bm\Phi^l \}_{l=1}^L, \bm\delta_a } &~ \left\| \bA^{\rm H} \bG \bold w^1 - \bg_a \odot \bm\delta_a \right\|_2^2 
		\\
		\mbox{s.t.}
		&~{\cal C}^G,{\cal C}^\phi, \left|\left[ \bm\delta_a \right]_i \right| = 1,~i=1,2,\dots,S_\theta S_\varphi,
	\end{split}
\end{equation}
where 
$\bA = \left[ \ba(\theta_1,\varphi_1), \dots, \ba(\theta_{S_\theta},\varphi_{S_\varphi}) \right] \in \mathbb{C}^{N\times S_\theta S_\varphi}$, $S_\theta$ and $S_\varphi$ denote the number of sampled points along the $\theta$ and $\varphi$ directions, respectively.
$\theta_{s}$ and $\varphi_{s}$ denote the angles of the $s$-th sampled point along $\theta$ and $\varphi$ directions, respectively.
$\bg_a$ is the given desired beam gain, which is set to 1 within the target angular range and 0 elsewhere.
In addition, to increase design flexibility, a variable $\bm\delta_a$ is introduced in the problem~\eqref{coded_codebook}, primarily for phase alignment of the beam, without affecting the beam gain~\cite{Zhang2025XL_RIS}.

However, in the above problem, $\bA$ is the steering matrix corresponding to the sampled points in the 2-D angle domain. 
Therefore, the computational complexity of solving this problem is proportional to $S_\theta^2 S_\varphi^2 L N$, which inevitably results in excessively high complexity.

\begin{figure}[t]	
	\centering \includegraphics[width=\linewidth]{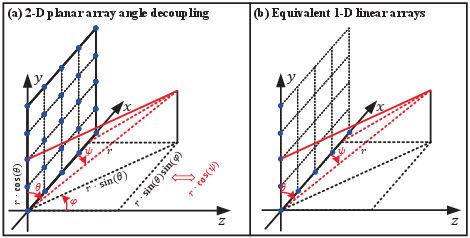}
	\vspace{-0.6cm}
	\caption{An illustration of two-dimensional angle decoupling and the equivalence between linear arrays and planar arrays.}
	\vspace{-0.6cm}
	\label{fig:2D_angle_decouple}
\end{figure}

Exploring a low-complexity codebook construction method in SIM systems is of great importance.
Therefore, we propose a TSCC approach: (i) design an ideal beamformer based on the desired beam pattern, and (ii) optimize the SIM phase shifts to approximate the ideal beamformer, thereby constructing the SIM codebook.
This strategy significantly reduces the complexity of SIM codebook construction.
Nonetheless, sampling over the 2-D angle domain is still required to achieve the desired codebook design, which continues to incur high computational complexity.
Therefore, we aim to perform independent sampling of each angle dimension, reducing the number of sampled points from $S_\theta^2 S_\varphi^2$ to $2S_\theta S_\varphi$, which can significantly lower the complexity.
However, since the angles $\varphi$ and $\theta$ are tightly coupled within the steering vector, sampling along a single angle dimension poses a considerable challenge.
To address this issue, we further propose an angle-domain decoupling method that enables the two angle dimensions to appear independently within the steering vector.
As shown in Fig.~\ref{fig:2D_angle_decouple}$\,$(a), through geometric theory, we have \vspace{-0.15cm}
\[
r\cdot {\rm sin}\left( \theta \right) {\rm sin}\left( \varphi \right) = r \cdot {\rm cos}\left( \psi \right).
\]\vspace{-0.6cm}

\noindent Based on this, we can convert the original steering vector $\ba(\theta,\varphi)$ to \vspace{-0.2cm}
\begin{equation*}
	\begin{split}
		\ba(\theta, \psi)
		=&\left[ 1, e^{-j\pi {\rm cos}(\psi)} , \dots,e^{-j\pi (N_1 - 1) {\rm cos}(\psi) }
		\right]^{\rm T} \\
		& \otimes \left[ 1, e^{-j\pi {\rm cos}(\theta)}, \dots,e^{-j\pi (N_2 - 1)  {\rm cos}(\theta)}
		\right]^{\rm T} \big/ \sqrt{N},
	\end{split}
\end{equation*}
\vspace{-0.15cm}

\noindent where $\varphi = {\rm arcsin}\left( \frac{ {\rm cos}(\psi) }{ {\rm sin}(\theta) } \right)$ and $\psi \in [0,\pi)$.

Further, we define $\nu \triangleq {\rm cos}(\theta) \in [-1,1]$ and $\vartheta \triangleq {\rm cos}(\psi) \in [-1,1]$.
Then, the steering vector can be rewritten as \vspace{-0.2cm}
\begin{equation*}
	\begin{split}
		\ba(\vartheta, \nu)
		=&\left[ 1, e^{-j\pi \vartheta} , \dots,e^{-j\pi (N_1 - 1) \vartheta }
		\right]^{\rm T} \\
		& \otimes \left[ 1, e^{-j\pi \nu}, \dots,e^{-j\pi (N_2 - 1)  \nu}
		\right]^{\rm T} \big/ \sqrt{N}.
	\end{split}
\end{equation*}

\vspace{-0.15cm}
Therefore, the planar array beamformer can be represented as the Kronecker product of the beamformers corresponding to the linear arrays along the $y$-axis and $x$-axis directions, as illustrated in Fig.~\ref{fig:2D_angle_decouple}$\,$(b).
Building on this observation, we propose a low-complexity TSCC method as follows.

\noindent $\blacklozenge$ {\bf{Step 1:}} We independently design the desired beamforming vectors for the $y$-axis and $x$-axis arrays. 
The corresponding optimization problems are formulated as follows \vspace{-0.15cm}
\begin{equation} \label{CC_step_1}
	\begin{cases}
		\min\limits_{ \bv_y, \left\{ \left|[\delta_y]_i\right|=1 \right\}_{i=1}^{S_y} } ~ \left\| \bA_y^{\rm H} \bv_y - \bg_y \odot \bm\delta_y \right\|_2^2 ,\\
		\min\limits_{ \bv_x, \left\{ \left|[\delta_x]_i\right|=1 \right\}_{i=1}^{S_x} } ~ \left\| \bA_x^{\rm H} \bv_x - \bg_x \odot \bm\delta_x \right\|_2^2 ,
	\end{cases}
\end{equation}
\vspace{-0.3cm}

\noindent where  \vspace{-0.25cm}
\[
\bA_y = \left[ \hat\ba(\nu_{1}), \dots, \hat\ba(\nu_{S_y}) \right],~\bA_x = \left[ \hat\ba(\vartheta_1), \dots, \hat\ba(\vartheta_{S_x}) \right],
\]
\[
\hat\ba(\hat\theta) = \left[ 1, e^{-j\pi \hat\theta} , e^{-j2\pi \hat\theta} , \dots,e^{-j\pi (\hat N - 1) \hat\theta }
\right]^{\rm T},
\]\vspace{-0.35cm}

\noindent $\hat N = N_2$ when $\hat\theta$ is $\vartheta$ and $\hat N = N_1$ when $\hat\theta$ is $\nu$.
$S_y$ and $S_x$ represent the numbers of sampled points within the range of $\vartheta$ and $\nu$, respectively. Here, we set $S_y = S_x = 180$.
$\bg_y$ and $\bg_x$ denote the desired beam gain vectors, while $\bm\delta_y$ and $\bm\delta_x$ are the corresponding desired phase shift vectors.
$\bv_y$ and $\bv_x$ denote the desired beamforming vectors for the $y$-axis and $x$-axis arrays, respectively. 

\noindent $\blacklozenge$ {\bf{Step 2:}} Our goal is to approximate the SIM wave-based beamforming vector to the desired beamforming vector obtained by solving problem~\eqref{CC_step_1}, thereby constructing the SIM codebook. Such process is formulated as follows \vspace{-0.15cm}
\begin{equation} \label{SIM_codebook_design}
	\begin{split}
		\min_{ \{ \bm\Phi^l \}_{l=1}^L } &~ \left\| \bG \bold w^1 - \bv_{x}^\star \otimes \bv_{y}^\star \right\|_2^2     \\
		\mbox{s.t.}	
		&~ {\cal C}^G: \bG = \bm\Phi^L \bW^L \bm\Phi^{L-1} \cdots \bm\Phi^2 \bW^2 \bm\Phi^1, \\
		&~ {\cal C}^{\phi}: \bPhi^l = \diag(\bm\phi^l),~|\phi_n^l| = 1,~l\in {\cal L},\,n \in {\cal N},
	\end{split}
\end{equation}
\vspace{-0.25cm}

\noindent where $\bv_{x}^\star$ and $\bv_{y}^\star$ denote the optimal solutions of the two problems in~\eqref{CC_step_1}, respectively.


{\it Remark 2:
Compared with directly performing joint optimization over the full multi-layer phase network and the phase vectors, the TSCC method preserves physical interpret ability while significantly reducing computational complexity and improving numerical stability. The minor performance differences mainly stem from the representational capacity of the SIM, which can be effectively mitigated by increasing the number of layers, applying a relax-and-project procedure, or using the two-step solution as initialization for limited joint fine-tuning, as proved in Appendix A. Moreover, the proposed TSCC method is more modular, amenable to parallel implementation and tuning.
}

\vspace{-0.3cm}
\subsection{Proposed Efficient Algorithm for TSCC}
In this section, we propose an efficient algorithm to address the codebook construction problem in the SIM-assisted communication system.
We first solve the problems~\eqref{CC_step_1} by using the relaxed Gerchberg-Saxton (GS) algorithm~\cite{Lu2024HierarchicalBeam,Chen2025CBT_RIS} to obtain the optimal desired beamforming vector $\bv = \bv_{y}^\star \otimes \bv_{x}^\star$.
Subsequently, the SIM codewords are derived by solving the following optimization problem
\begin{equation} \label{SIM_codebook_problem}
	\begin{split}
		\min_{ \{ \bm\Phi^l \}_{l=1}^L } ~ \left\| \bG \bold w^1 - \bv \right\|_2^2 ~~~~ \mbox{s.t.}
		~ {\cal C}^G,~ {\cal C}^{\phi}.
	\end{split}
\end{equation}

This problem is highly challenging to solve because the $L$ layers of phase shift matrices are coupled and each is subject to non-convex unit-modulus constraints.
To address this problem, we first apply an alternating optimization (AO) algorithm to decouple the phase shift matrices across different layers.
Therefore, the problem~\eqref{SIM_codebook_problem} can be efficiently solved using the following update procedure.
\begin{equation}\label{AO_SIM}
	\begin{split}
		\bPhi^{1,\ell+1} &= ~\mathop{\rm argmin}\limits_{\bPhi^1}~ \left\| \bG \bold w^1 - \bv \right\|_2^2  \\
		&\mbox{s.t.}
		~ \bG = \bm\Phi^{L,\ell} \bW^L \cdots \bm\Phi^{2,\ell} \bW^{2} \bm\Phi^1,~ {\cal C}^{\phi}, 
		\\
		\bPhi^{2,\ell+1} &= ~\mathop{\rm argmin}\limits_{\bPhi^2}~ \left\| \bG \bold w^1 - \bv \right\|_2^2  \\
		&\mbox{s.t.}
		~ \bG = \bm\Phi^{L,\ell} \bW^L \cdots \bm\Phi^{2} \bW^{2} \bm\Phi^{1,\ell+1},~ {\cal C}^{\phi}, 
		\\
		\bm\cdots~\bm\cdots   \\
		\bPhi^{L,\ell+1} &= ~\mathop{\rm argmin}\limits_{\bPhi^L}~ \left\| \bG \bold w^1 - \bv \right\|_2^2  \\
		&\mbox{s.t.}
		~ \bG = \bm\Phi^{L} \bW^L \bm\Phi^{L-1,\ell+1} \cdots \bm\Phi^{1,\ell+1},~ {\cal C}^{\phi}. 
	\end{split}
\end{equation}
\begin{algorithm}[t!]
	\caption{ AO-PDMM Algorithm for Problem~\eqref{SIM_codebook_problem} }
	\begin{algorithmic}[1]
		\STATE  {\bf Input:}  initialize $\bm\phi^{l,0}$, $\mu = 0.1$, $\epsilon=2$, $L_{\rm num}=4$; $\ell=0$. 
		\REPEAT
		\STATE {\bf{for $l$ from $1$ to $L$:}}
		\STATE $~$ {\bf{repeat}}
		\STATE $~\quad$  $\Pi_{{\cal C}_{\phi}}(\hat{\bm\phi}^{l,i}) = e^{j\angle(\hat{\bm\phi}^{l,i})}$;
		\STATE $~\quad$  $\hat{\bm\phi}^{l,i+1} = \left( \bC^{l,{\rm H}} \bC^l + \mu \bI \right)^{-1} ( \bC^{l,{\rm H}}\bv + \mu \Pi_{{\cal C}^{\phi}}(\hat{\bm\phi}^{l,i}) )$;
		\STATE $~\quad$ $\mu = \epsilon\cdot\mu$ every $L_{\rm num}$ iterations;
		\STATE $~\quad$ $i = i + 1$;
		\STATE $~$ {\bf{until}} $\|\hat{\bm\phi}^{l,i+1} - \Pi_{{\cal C}^{\phi}}(\hat{\bm\phi}^{l,i}) \|_2 \leq 10^{-4}$;
		\STATE $~$ $\bm\phi^{l,\ell+1} = \hat{\bm\phi}^{l,i+1}$;
		\STATE {\bf{end}}
		\STATE \bf{update} $\bG^{\ell+1} = \bm\Phi^{L,\ell+1} \bW^L \bm\Phi^{L-1,\ell+1}  \cdots \bW^2 \bm\Phi^{1,\ell+1}$; 
		\STATE $\ell= \ell+1$;
		\UNTIL $\left| \left\| \bG^{\ell+1}\bold w^1 - \bv \right\|_2^2 - \left\| \bG^{\ell}\bold w^1 - \bv \right\|_2^2 \right| \leq 10^{-5} $.
	\end{algorithmic}
\end{algorithm} \vspace{-0.2cm}
\begin{figure}[t]	
	\centering \includegraphics[width=\linewidth]{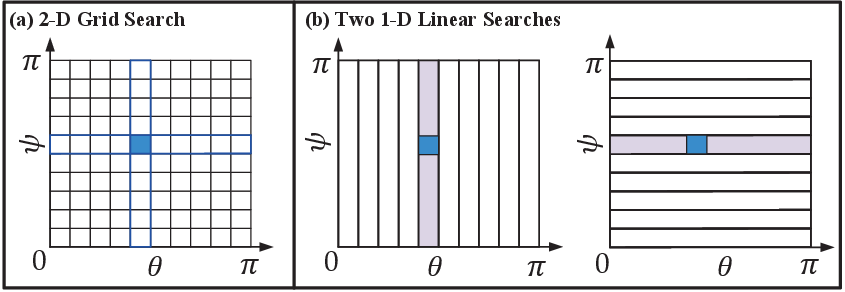}
	\vspace{-0.65cm}
	\caption{Different beam training strategies. (a) Traditional 2-D grid search; (b) Proposed 1-D linear search.}
	\vspace{-0.6cm}
	\label{fig:Exhuastive_Beam_Training}
\end{figure}

To explicitly express the variables in each subproblem, 
we then reformulate the wave-based beamforming vector $\bG \bold w^1$ as $\bC^l \bm\phi^l$ for any $l\in {\cal L}$, where \vspace{-0.1cm}
\[
\bC^1 = \bm\Phi^L \bW^L \cdots \bW^2 \diag(\bold w^1),
\bC^L = \diag(\bW^L \cdots \bm\Phi^1\bold w^1),
\]
\[
\bC^l = \bm\Phi^L\bW^L \cdots \bW^{l+1} \diag(\bW^l \cdots \bW^2 \bm\Phi^1 \bold w^1),l\in[2,L-1].
\]

Therefore, the subproblem of \eqref{AO_SIM} with respect to $\bm\phi^l$ can be
equivalently reformulated as \vspace{-0.15cm}
\begin{equation} \label{CC_phi}
	\begin{split}
		\min_{ \bm\phi^l } ~ \left\| \bC^l\bm\phi^l - \bv \right\|_2^2 ~~~	\mbox{s.t.}	
		~ {\cal C}^{\phi}: |\phi_n^l| = 1,~n \in {\cal N}.
	\end{split}
\end{equation}

\vspace{-0.25cm}
The above problem remains non-convex due to the constant-modulus constraints.
To tackle this, we transform it into an unconstrained optimization problem using the proximal distance (PD) algorithm~\cite{Keys2019PDA,Zhang2024PDA}, formulated as follows \vspace{-0.15cm}
\begin{equation} \label{PDA_CC}
	\begin{split}
		\min_{ \bm\phi^l } ~ \left\| \bC^l\bm\phi^l - \bv \right\|_2^2 + \mu \cdot {\rm dist}^2\left( \bm\phi^l, {\cal C}^{\phi} \right),
	\end{split}
\end{equation}
\vspace{-0.5cm}

\noindent where ${\rm dist}\left( \bm\phi^l, {\cal C}^{\phi} \right)$ denotes the Euclidean distance between the variable $\bm\phi^l$ and the space constrained by ${\cal C}^{\phi}$.
Problems~\eqref{CC_phi} and~\eqref{PDA_CC} share the same optimal solution as $\mu \rightarrow \infty$.

However, the explicit form of ${\rm dist}\left( \bm\phi^l, {\cal C}^{\phi} \right)$ is difficult to obtain. Therefore, by the majorization-minimization (MM) algorithm~\cite{Sun2017MM,Landeros2023MM}, we can majorize the distance as \vspace{-0.15cm}
\begin{equation}
	\begin{split}
		{\rm dist}\left( \bm\phi^l, {\cal C}^{\phi} \right) \leq \|\bm\phi^l - \hat{\bm\phi} \|_2,
	\end{split}
\end{equation}\\
\vspace{-1cm}

\noindent where $\hat{\bm\phi}$ is an arbitrary point in the space constrained by ${\cal C}^{\phi}$.

Therefore, by the PDMM algorithm, the problem~\eqref{PDA_CC} is equivalent to the following explicit problem
\begin{equation} \label{PDMM_CC}
	\begin{split}
		\min_{ \bm\phi^l } ~ \left\| \bC^l\bm\phi^l - \bv \right\|_2^2 + \mu \cdot \|\bm\phi^l - \Pi_{{\cal C}^{\phi}}(\hat{\bm\phi}^l) \|_2^2,
	\end{split}
\end{equation}
where $\hat{\bm\phi}^l$ is the solution in the previous iteration and 
the projection $\Pi_{{\cal C}^{\phi}}(\hat{\bm\phi}^l) = \arg\min_{\bz \in \{ \bz\in \mathbb{C}^{N}: |z_n|=1 \}} \|\bz - \hat{\bm\phi}^l \|_2^2$ maps $\hat{\bm\phi}^l$ onto the feasible set ${\cal C}^\phi$, ensuring that the solution satisfies the constant-modulus hardware constraints of the meta-atoms.
Problem~\eqref{PDMM_CC} can converge the same KKT point as the corresponding subproblems in~\eqref{AO_SIM}, as proved in Appendix~B.

Therefore, by equating the derivative of the objective function with respect to $\bm\phi^l$ to zero, the closed-form solution to optimization problem~\eqref{PDMM_CC} can be derived as follows.\vspace{-0.1cm}
\begin{equation}
	\begin{split}
		(\bm\phi^l)^{\star} = \left( \bC^{l,{\rm H}} \bC^l + \mu\cdot \bI \right)^{-1} \left( \bC^{l,{\rm H}}\bv + \mu \cdot \Pi_{{\cal C}^{\phi}} (\hat{\bm\phi}^l) \right),
	\end{split}
\end{equation}
where for $n\in {\cal N}$, we have \vspace{-0.1cm}
\begin{equation} 
	\left[\Pi_{{\cal C}^{\phi}}(\hat{\bm\phi}^l)\right]_n = 
	\begin{cases}
		\hat\phi_n^l,~\mbox{if}~ |\hat\phi_n^l|=1, \\
		\frac{\hat\phi_n^l}{|\hat\phi_n^l|}, ~\mbox{otherwise}.
	\end{cases}
\end{equation}
The AO-PDMM algorithm is summarized in {\bf{Algorithm~1}}.


\begin{figure*}[t]	
	\centering \includegraphics[width=\linewidth]{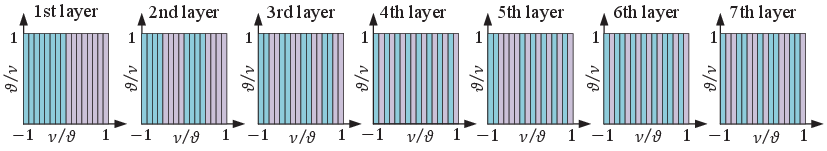}
	\vspace{-0.75cm}
	\caption{Angle coverage range corresponding to the ideal beam pattern. }
	\vspace{-0.5cm}
	\label{fig:HanmmingCodeBeamTraining}
\end{figure*}

\vspace{-0.15cm}
\section{Two-Stage Coded-Sliding Beam Training}
In this section, we propose an efficient TSCSBT method to acquire CSI.
In the first stage, we introduce the CBT method with the main idea illustrated using the (7,4) Hamming code as an example.
In the second stage, we propose a sliding sampling strategy based on the beam-width characteristics to enable high-resolution beam training.
Furthermore, based on the one-dimensional (1-D) line-search framework, we further propose a MP-TSBT scheme to support multi-path users.

\vspace{-0.35cm}
\subsection{Coded Beam Training} \vspace{-0.1cm}
Due to the planar array structure of SIM, the BT scheme in SIM systems is designed jointly over both angle dimensions, which inevitably leads to grid-based interleaving and increased complexity, as illustrated in Fig.~\ref{fig:Exhuastive_Beam_Training}$\,$(a).
To address this issue, we aim to decompose the 2-D angle space into separate 1-D angle space, enabling the BT design to be performed along a single angle direction, as shown in Fig.~\ref{fig:Exhuastive_Beam_Training}$\,$(b).
Fortunately, with the proposed 2-D angle decoupling method, the coupled angles can be effectively separated, allowing for simplified 1-D processing. 
Furthermore, the proposed TSCC method facilitates the efficient construction of the SIM codebook.
Therefore, we decompose the HBT design in the 2-D angular domain into separate 1-D angular domains to enable efficient training.
Moreover, to improve the training accuracy of HBT, we introduce the CBT scheme and exploit its processing procedure by embedding the (7,4) Hamming code into a 4-layer BT structure as an example.
Specifically, the CBT is divided into two stages, i.e., beam coding and beam decoding.

\noindent $\blacklozenge$ {\bf{Beam Coding:}}
For each angle dimension, the each layer of HBT corresponds to a different feedback symbol. 
Specifically, we divide the angle range $[-1,1]$ into 16 uniform regions, each corresponding to a feedback symbol given as
\begin{equation*}
	\bF = \left[
	\begin{array}{c}
		0 ~~ 0 ~~ 0 ~~ 0 ~~ 0 ~~ 0 ~~ 0 ~~ 0 ~~ 1 ~~ 1 ~~ 1 ~~ 1 ~~ 1 ~~ 1 ~~ 1 ~~ 1  \\
		0 ~~ 0 ~~ 0 ~~ 0 ~~ 1 ~~ 1 ~~ 1 ~~ 1 ~~ 0 ~~ 0 ~~ 0 ~~ 0 ~~ 1 ~~ 1 ~~ 1 ~~ 1  \\
		0 ~~ 0 ~~ 1 ~~ 1 ~~ 0 ~~ 0 ~~ 1 ~~ 1 ~~ 0 ~~ 0 ~~ 1 ~~ 1 ~~ 0 ~~ 0 ~~ 1 ~~ 1  \\
		0 ~~ 1 ~~ 0 ~~ 1 ~~ 0 ~~ 1 ~~ 0 ~~ 1 ~~ 0 ~~ 1 ~~ 0 ~~ 1 ~~ 0 ~~ 1 ~~ 0 ~~ 1
	\end{array}
	\right],
\end{equation*}
where
the first through fourth rows of matrix $\bF$ correspond to the beam patterns of the first to fourth layers, respectively, in Fig.~\ref{fig:HanmmingCodeBeamTraining}, referred to as the {\it information layers}.

To enhance the beam training accuracy, we employ the (7,4) Hamming code to encode the beams, thereby introducing self-correction capability and generating three additional {\it check layers}.
Based on this design, the $s$-th coded bitstream $\bx_s$, corresponding to the $s$-th spatial direction, can be expressed as follows
\begin{equation}
	\begin{split}
		\bx_s = {\rm mod}\left( \bm f_s \bE,2 \right) \in \left\{ 0,1 \right\}^7,~s=1,2,\dots,16,
	\end{split}
\end{equation}
where $\bm f_s$ is the $s$-th column of matrix $\bF$ and $\bE$ is the generator as follows  \vspace{-0.15cm}
\begin{equation*}
	\bE = \left[
	\begin{array}{c|c}
		  1~~0~~0~~0 & 1~~1~~1 \\
		  0~~1~~0~~0 & 1~~1~~0 \\
		  0~~0~~1~~0 & 1~~0~~1 \\
		  0~~0~~0~~1 & 0~~1~~1
	\end{array}
	\right].
\end{equation*}

 \vspace{-0.15cm}
As a result, we can obtain the corresponding matrix of feedback symbols for {\it check layers} as \vspace{-0.1cm}
\begin{equation*}
	\bJ = \left[
	\begin{array}{c}
		0 ~~ 0 ~~ 1 ~~ 1 ~~ 1 ~~ 1 ~~ 0 ~~ 1 ~~ 0 ~~ 1 ~~ 0 ~~ 1 ~~ 0 ~~ 1 ~~ 0 ~~ 1  \\
		0 ~~ 1 ~~ 0 ~~ 0 ~~ 1 ~~ 0 ~~ 1 ~~ 0 ~~ 1 ~~ 0 ~~ 1 ~~ 0 ~~ 0 ~~ 1 ~~ 0 ~~ 1  \\
		0 ~~ 1 ~~ 1 ~~ 0 ~~ 0 ~~ 1 ~~ 1 ~~ 0 ~~ 1 ~~ 0 ~~ 0 ~~ 1 ~~ 1 ~~ 0 ~~ 0 ~~ 1
	\end{array}
	\right].
\end{equation*}

\vspace{-0.1cm}
Then, based on the matrix $\bJ$, we design three check beam patterns, as illustrated in the last three subfigures of Fig.~\ref{fig:HanmmingCodeBeamTraining}.
This design process is carried out separately for both the $\nu$ and $\vartheta$ directions to ensure that the self-correction of feedback symbols can be achieved in both angular dimensions.
Once the desired beam patterns are determined, the corresponding codebook can be constructed using the proposed TSCC method.

\noindent $\blacklozenge$ {\bf{Beam Decoding:}}
Through the beam training, we obtain two feedback symbol vectors corresponding to the two angular dimensions, respectively.
For ease of interpretation, we uniformly denote the received feedback symbol vector as $\bm f \in \mathbb{C}^{7 \times 1}$.
Our objective is to recover the spatial direction index from $\bm f$.
To this end, we first define the parity-check matrix $\bH$ and the error pattern matrix $\bE_d$ as \vspace{-0.1cm}
\begin{equation*}
	\bH = \left[
	\begin{array}{c|c}
		1~~1~~1~~0 & 1~~0~~0 \\
		1~~1~~0~~1 & 0~~1~~0 \\
		1~~0~~1~~1 & 0~~0~~1
	\end{array}
	\right],
\end{equation*}
\begin{equation*}
	\bE_d = \left[
	\begin{array}{c}
		1 ~~ 1 ~~ 1 ~~ 0 ~~ 1 ~~ 0 ~~ 0 ~~ 0   \\
		1 ~~ 1 ~~ 0 ~~ 1 ~~ 0 ~~ 1 ~~ 0 ~~ 0   \\
		1 ~~ 0 ~~ 1 ~~ 1 ~~ 0 ~~ 0 ~~ 1 ~~ 0 
	\end{array}
	\right].
\end{equation*}

\vspace{-0.1cm}
By computing $\bc = {\rm mod}\left( \bH\bm f,2 \right)$, we obtain the decoding vector $\bc\in \mathbb{C}^{3\times 1}$.
Next, the Hamming distance between $\bc$ and each column of the matrix $\bE_d$ is calculated.
The column index corresponding to the minimum Hamming distance indicates the position of the erroneous bit in $\bm f$.
Exceptionally, if the minimum Hamming distance corresponds to column index 8, it implies that no bit error has occurred in $\bm f$.

For example, suppose the true feedback symbol vector is $[0~1~0~1~1~0~1]^{\rm T}$, while the received vector is $\bm f = [0~1~0~0~1~0~1]^{\rm T}$.
Computing $\bc = {\rm mod}\left( \bH\bm f,2 \right) = [0~1~1]^{\rm T}$, and then evaluating the Hamming distances between $\bc$ and each column of $\bE_d$ reveals that the fourth bit of $\bm f$ is erroneous.
Correcting this bit yields $[0~1~0~1~1~0~1]^{\rm T}$, matching the true symbol vector. 
This demonstrates the 1-bit error-correcting capability of CBT enabled by the (7,4) Hamming code beam encoding and decoding.

\begin{figure}[t]	
	\centering \includegraphics[width=0.9\linewidth]{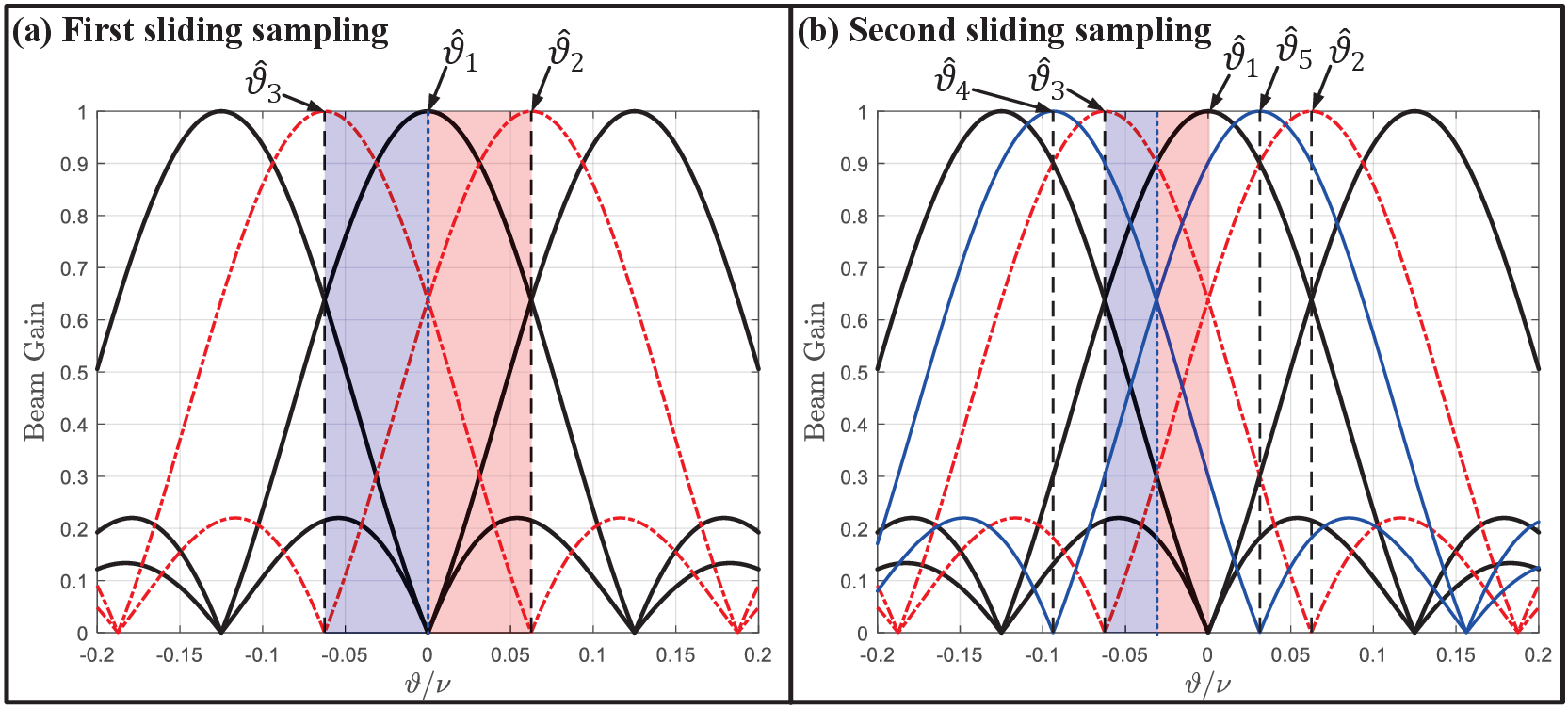}
	\vspace{-0.35cm}
	\caption{Illustration of the sliding sampling model in sliding beam training.}
	\vspace{-0.55cm}
	\label{fig:Sliding_beam_training}
\end{figure}
\begin{figure}[t]
	\centering
	\vspace{-0.32cm}
	\subfloat[Proposed TLS-BT]{\includegraphics[width=0.45\columnwidth]{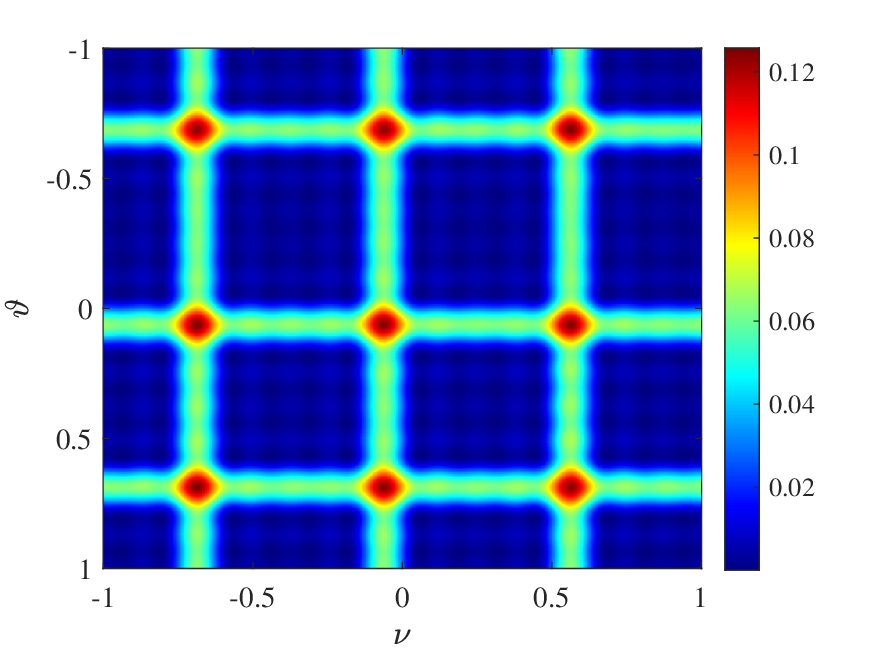}%
		\label{Multi_path_BP_example}}
	\hfil
	\subfloat[Candidate angle sweeping ]{\includegraphics[width=0.45\columnwidth]{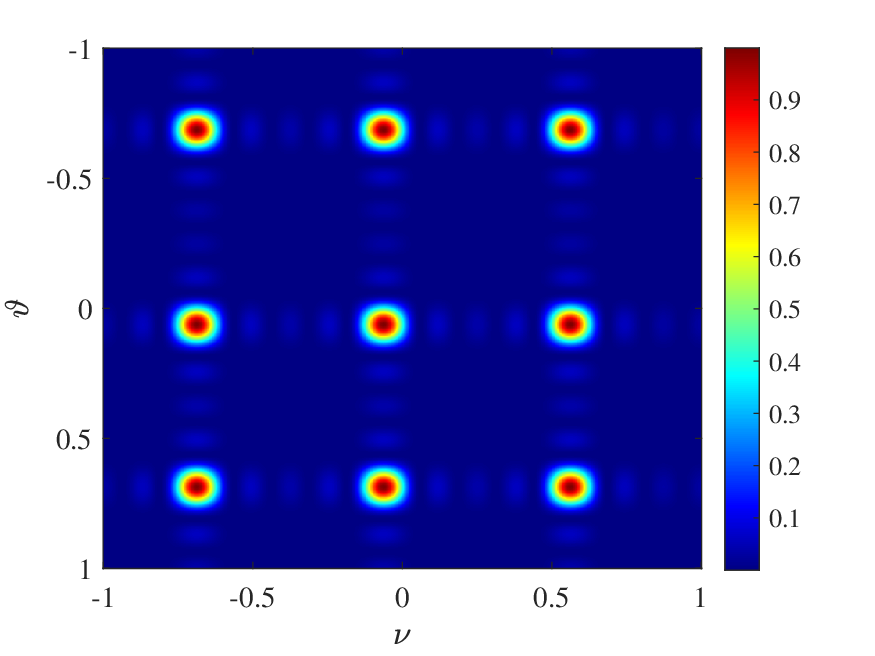}%
		\label{Multi_path_BP_candidate_angle_example}}
	\vspace{-0.1cm}
	\caption{Search procedure for the proposed MP-TSBT: (a) candidate directions of each path identified using the proposed TLS-BT method, and (b) scanning performed around the obtained candidate angles.}
	\vspace{-0.55cm}
	\label{fig:Multi_path_BP}
\end{figure}

\vspace{-0.55cm}
\subsection{Sliding Beam Training} \vspace{-0.1cm}
Through the aforementioned coded beam training (CBT) method, the beam coverage of the sampling points can be accurately determined. 
However, the spatial angular resolution is inherently constrained by the number of array elements. 
To further improve the angular resolution, we propose a strategy that refines the sampling points by sliding their positions. 
The SIM exhibits a strong capability to approximate the steering vector associated with the typical user located at $(\vartheta_u,\nu_u)$. 
Therefore, in the angular domain corresponding to $\vartheta$, the peak beam gain occurs at the angle $\vartheta_u$ associated with the typical user.
The angular distance between the first nulls on both sides of the main lobe is defined as the beam-width, which is given by $\frac{4}{N_2}$~\cite{Cui2022PolarCodebook}.
Similarly, in the $\nu$-domain, the beam-width is given by $\frac{4}{N_2}$.
To ensure robustness against noise and achieve full coverage of the angular domain, we perform double sampling in the angle range~\cite{Cui2022PolarCodebook}, given by $\frac{2n - N_1 + 1}{N_1}$, where $n=0,1,\dots,N_1-1$.

After the CBT method identifies the sampling angle closest to the user, a sliding sampling procedure is conducted around that angle and its neighboring directions.
As illustrated in Fig.~\ref{fig:Sliding_beam_training}$\,$(a), assume that the sampling angle obtained by CBT corresponds to the peak of the central black line, denoted by $\hat\vartheta_1$.
We then shift the sampling points $\hat\vartheta_1$ and $\hat\vartheta_1 + \frac{2}{N_1}$ leftward by a distance of $\frac{1}{N_1}$, resulting in two new beams corresponding to the angles $\hat\vartheta_3$ and  $\hat\vartheta_2$ for beam training.
As shown in Fig.~\ref{fig:Sliding_beam_training}$\,$(b), if the beam corresponding to $\hat\vartheta_3$ provides a higher beam gain to the user than that corresponding to $\hat\vartheta_2$, the sampling points $\hat\vartheta_3$ and $\hat\vartheta_2$ are further shifted leftward by $\frac{0.5}{N_1}$, yielding two additional beams corresponding to the angles $\hat\vartheta_4$ and $\hat\vartheta_5$ for beam training.
If the beam at $\hat\vartheta_4$ delivers the higher gain to the user, the final user angle obtained through the sliding beam training procedure can be determined as $\frac{\hat\vartheta_3 + \frac{\hat\vartheta_3 + \hat\vartheta_1}{2}}{2}$.
In summary, by leveraging the beam-width characteristics, adjacent sampling points can be sequentially slid, enabling a progressively finer angular resolution.

\vspace{-0.5cm}
\subsection{Multi-Path Two-Stage Beam Training}  \vspace{-0.1cm}
Based on the proposed independent angular domain line-search framework, we develop a MP-TSBT scheme to effectively capture user’s multi-path information.
As illustrated in Fig.~\ref{fig:Exhuastive_Beam_Training}, the two angular domains are decoupled during the beam training process, allowing independent two line-search beam training (TLS-BT) operations. 
This approach significantly reduces the training overhead from $N_1\times N_2$ to $N_1 + N_2$.
Therefore, in the first stage of the proposed MP-TSBT scheme, beam training is performed separately in the two angular domains. 
Each domain then feeds back the angles corresponding to the top $P$ received power values that exceed the predefined threshold $P_{\rm th}$, in order to obtain the candidate multi-path angles.
In the second stage, beam training is carried out within the set of these candidate angles, enabling efficient acquisition of user multi-path information with substantially reduced training overhead.

As an illustrative example, we consider the case of $P=3$.
In the first stage, the TLS-BT procedure is conducted to acquire the corresponding angles in the two angular domains, yielding nine candidate angles, as shown in Fig.~\ref{fig:Multi_path_BP}$\,$(a).
Subsequently, as illustrated in Fig.~\ref{fig:Multi_path_BP}$\,$(b), beam sweeping is performed over these candidate angles to identify the final $P$ multi-path components.

\vspace{-0.45cm}
\section{Proposed VD-BSUM Algorithm for QoS-Constrained SRM} \vspace{-0.1cm}
Using the proposed beam training method, we can efficiently acquire the users’ CSI, which can then be used for the beamforming design in multiuser communications.
Therefore, in this section, we develop a low-complexity VD-BSUM algorithm to efficiently solve the SRM problem with QoS constraints, enabling effective management of inter-user interference.
Firstly, we transform the problem~\eqref{WSRM_QoS} by the weighted minimum-mean-square-error (WMMSE) method~\cite{shi2011wmmse} into \vspace{-0.1cm}
\begin{equation}
	\begin{split}
		\min_{ \{ \bm\phi^l \}_{l=1}^L, \bp, \bu, \bm\zeta } {\cal F}(\bp, \bG, \bu, \bm\zeta)~{\rm s.t.}~{\cal C}^G,{\cal C}^\phi,{\cal C}^p,{\cal C}_k^{R},k\in{\cal K},
	\end{split}
\end{equation}
\vspace{-0.45cm}

\noindent where $\bu = \left[ u_1,u_2,\dots,u_K \right]^{\rm T}$ and $\bm\zeta = \left[ \zeta_1,\zeta_2,\dots,\zeta_K \right]^{\rm T}$ denote the introduced auxiliary variables, and \vspace{-0.1cm}
\[
{\cal F}(\bp, \bG, \bu, \bm\zeta)  =  \sum_{k=1}^{K} \zeta_k \cdot \left( {\cal G}_k + 1 \right)
 - \sum_{k=1}^{K} {\rm log}\,(\zeta_k),
\]
\[
{\cal G}_k =   |u_k|^2 \left( \sum_{i=1}^{K} \left| \bh_k^{\rm H} \bG \bw_i^1 \right|^2 p_i^2 + \sigma_k^2 \right) - 2\Re\left\{ u_k^* \bh_k^{\rm H} \bG \bw_k^1 p_k \right\},
\]
where $\sqrt{\overline p_k}$ is replaced by $p_k$ for efficient optimization.

Subsequently, we employ the BSUM algorithm to solve it, which follows the updates rule
\begin{subequations}\label{AO_SRM}
	\begin{align}
		\bu^{\ell+1} = &~\mathop{\rm argmin}\limits_{\bu}~ {\cal F}(\bp^{\ell}, \bG^\ell, \bu, \bm\zeta^{\ell}), \label{AO_SRM_1}
		\\
		\bm\zeta^{\ell+1} = &~\mathop{\rm argmin}\limits_{\bm\zeta}~ {\cal F}(\bp^{\ell},  \bG^\ell, \bu^{\ell+1}, \bm\zeta), \label{AO_SRM_2}
		\\
		\bp^{\ell+1} = &~\mathop{\rm argmin}\limits_{\bp \in {\cal C}^{p}\cap {\cal C}_{k}^R}~ {\cal F}(\bp, \bG^\ell, \bu^{\ell+1}, \bm\zeta^{\ell+1}), \label{AO_SRM_3}
		\\
		\bm\phi^{l,\ell+1} = &~\mathop{\rm argmin}\limits_{\bm\phi^l \in {\cal C}^G \cap {\cal C}^\phi\cap {\cal C}_{k}^R}  {\cal F}(\bp^{\ell+1},  \{ \bm\phi^{i,\ell} \}_{i\neq l}^{L}, \bu^{\ell+1}, \bm\zeta^{\ell+1}). \label{AO_SRM_4}
	\end{align}
\end{subequations}

According to subproblems~\eqref{AO_SRM_1} and~\eqref{AO_SRM_2}, the optimal solutions of $u_k$ and $\zeta_k$ can be derived as \vspace{-0.1cm}
\[
u_k^\star = \frac{ \bh_k^{\rm H} \bG \bw_k^1  p_k }{\sum_{i=1}^{K}| \bh_k^{\rm H} \bG \bw_i^1  |^2  p_i^2 + \sigma_k^2 },~~
\zeta_k^\star = \left( 1 - u_k^* \bh_k^{\rm H} \bG \bw_k^1  p_k \right)^{-1},
\]
for any $k \in {\cal K}$.

\vspace{-0.3cm}
\subsection{Optimize $\bp$}
The problem~\eqref{AO_SRM_3} with respect to $\bp$ can be rewritten as \vspace{-0.15cm}
\begin{equation} \label{SRM_QoS_P}
	\begin{split}
		\min_{ \bp } ~ \sum_{i=1}^{K} \left[ a_i p_i^2 - 2 b_i p_i  \right]~
		\mbox{s.t.}	
		~{\cal C}^p,~{\cal C}_k^R,~ k\in {\cal K},
	\end{split}
\end{equation}
\vspace{-0.15cm}
 
\noindent where $a_i = \Re\left\{ \bw_i^{1,{\rm H}} \bG^{\rm H} \left( \sum_{k=1}^{K}  \zeta_k |u_k|^2 \bh_k\bh_k^{\rm H} \right) \bG \bw_i^1  \right\}$ and $b_i = \Re\left\{ \zeta_i u_i^* \bh_i^{\rm H} \bG \bw_i^1 \right\}$.

By PDMM algorithm, the problem~\eqref{SRM_QoS_P} can further be transformed into
\begin{equation*} \label{WSRM_QoS_P_PDA}
	\begin{split}
		\min_{ \bp } \sum_{i=1}^{K} \left[ a_i p_i^2 - 2 b_i p_i + \mu \| \bp - \Pi_{{\cal C}_{i}^R}(\hat{\bp}) \|_2^2  \right] + \mu \| \bp - \Pi_{{\cal C}^p} (\hat{\bp}) \|_2^2,
	\end{split}
\end{equation*}
where $\hat{\bp}$ is a projection point that takes the value of the optimal solution of the previous iteration.
It is noteworthy that the penalty terms associated with different constraints are of the same order of magnitude. 
To ensure a balanced penalization and avoid bias toward any particular constraint, a unified penalty parameter $\mu$ is employed across all constraints. 
Moreover, as the penalty factor increases, the solution of the penalized problem asymptotically approaches the feasible solution of the original constrained problem~\cite{Bertsekas1999NLP}. 
Consequently, the use of a common penalty parameter does not compromise the convergence to the final solution. The closed-form solution of the above problem is derived as
\begin{equation} \label{closed_form_p}
	\begin{split}
		\bp = \left( \hat\bA + \mu(K+1)\bI_K \right)^{-1}\left( \hat\bb + \mu \Pi_{{\cal C}^{p} } (\hat{\bp}) + \mu \sum_{i=1}^{K}\Pi_{{\cal C}_{i}^R}(\hat{\bp}) \right),
	\end{split}
\end{equation}
where $\hat b_i = b_i$, and if $i=k$ then $\hat A_{i,k} = a_i$ otherwise $\hat A_{i,k} = 0$.

Next, we further propose efficient algorithms to obtain the projection solutions $\Pi_{{\cal C}^{p} } (\hat{\bp})$ and $\Pi_{{\cal C}_{i}^R}(\hat{\bp})$ in the above closed-form solutions.

\subsubsection{Solve $\Pi_{{\cal C}^{p} } (\hat{\bp})$}
The projection solution $\Pi_{{\cal C}^{p} } (\hat{\bp})$ can be obtained by solving the following problem
\begin{equation} \label{WSRM_QoS_proj_Cp}
	\begin{split}
		\min_{ \bp } ~ \left\| \bp - \hat\bp \right\|_2^2 ~~
		\mbox{s.t.}	
		~\sum_{i=1}^{K} p_i^2 \leq P_{\rm max},~p_i\geq0, i \in {\cal K}.
	\end{split}
\end{equation}

The Lagrangian function of the problem~\eqref{WSRM_QoS_proj_Cp} is given by
\[
{\cal L}_{p} = \left\| \bp - \hat\bp \right\|_2^2 + \rho \left( \sum_{i=1}^{K} p_i^2 - P_{\rm max} \right) - \gamma_i p_i,
\]
where $\rho \geq 0$ and $\gamma_i\geq 0$ denote the dual variables.

The Karush–Kuhn-Tucker (KKT) conditions are given by
\begin{equation*}
	\begin{split}
		\frac{\partial {\cal L}_{p}}{\partial p_i} = (2 + 2\rho) p_i - 2 \hat p_i - \gamma_i
		= 0;
	\end{split}
\end{equation*}
\begin{equation*}
	\begin{split}
		\gamma_i\geq 0;~ \gamma_i p_i = 0;~ \sum_{i=1}^{K} p_i^2 \leq P_{\rm max};~ \rho \left( \sum_{i=1}^{K} p_i^2 - P_{\rm max} \right)=0.
	\end{split}
\end{equation*}

According to the complementary slackness, we know that $\gamma_i = 0$ if $p_i>0$ and otherwise $\gamma_i >0$ if $p_i = 0$.
If $p_i>0$, according to the first line of the above KKT conditions, we have $p_i = \frac{\hat p_i}{1 + \rho}$ with $\hat p_i > 0$ and otherwise $p_i = 0$ if $\hat p_i \leq 0$.
Therefore, we have the optimal solution as
\[
p_i = {\rm max}\left( 0, \frac{\hat p_i}{1 + \rho} \right),~i \in {\cal K},
\]
where $\rho$ can be found by the 1-D bisection search such that $\sum_{i=1}^{K} \left( {\rm max}\left( 0, \frac{\hat p_i}{1 + \rho} \right) \right)^2 = P_{\rm max} $.

\subsubsection{Solve $\Pi_{{\cal C}_{i}^R}(\hat{\bp})$}
The projection solution $\Pi_{{\cal C}_{i}^R}(\hat{\bp})$ can be obtained by solving the following problem
\begin{equation} \label{WSRM_QoS_proj_Ci}
	\begin{split}
		\min_{ \bp } \left\| \bp - \hat\bp \right\|_2^2 ~
		\mbox{s.t.}	
		\frac{ | \bh_k^{\rm H} \bG \bw_k^1  |^2 p_k^2 }{\sum_{i=1,i\neq k}^{K}| \bh_k^{\rm H} \bG \bw_i^1  |^2  p_i^2 + \sigma_k^2 } \geq \gamma_{\rm th}.
	\end{split}
\end{equation}

However, the problem~\eqref{WSRM_QoS_proj_Ci} remains intractable due to the non-convex constraint.
Therefore, to efficiently solve it, we introduce a new variable $\bz = \bp$ to decouple the complicated constraint as follows
\begin{equation} \label{SRM_QoS_proj_Ci}
	\begin{split}
		\min_{ \bp,\bz } &~ \left\| \bp - \hat\bp \right\|_2^2    \\
		\mbox{s.t.}	
		&~ \bz = \bp, 
		~ {\cal C}^{pz}: \frac{ a_k p_k z_k }{\sum_{i=1,i\neq k}^{K} a_i p_i z_i + \sigma_k^2 } \geq \gamma_{\rm th},
	\end{split}
\end{equation}
where $a_i = | \bh_k^{\rm H} \bG \bw_i^1  |^2$ for given $k$.

Then, we use the IPDD algorithm~\cite{Shi2020PDD} to solve the problem~\eqref{SRM_QoS_proj_Ci} by the following updates
\begin{subequations}\label{IPDD_p_Ci}
	\begin{align}
		\bp^{t+1} = &~\mathop{\rm argmin}\limits_{\bp \in {\cal C}^{pz}} \left\| \bp - \hat\bp \right\|_2^2 + \frac{1}{2\eta^t} \left\| \bz^{t} - \bp + \eta^t \bm\omega^{t} \right\|_2^2, \label{IPDD_p_1}
		\\
		\bz^{t+1} = &~\mathop{\rm argmin}\limits_{\bz \in {\cal C}^{pz}} \left\| \bz - \bp^{t+1} + \eta^t \bm\omega^{t} \right\|_2^2, \label{IPDD_p_2} \\
		\bm\omega^{t+1} =&~\bm\omega^{t} + (\bz^{t+1} - \bp^{t+1}) \big/ \eta^t, \label{IPDD_p_3}
	\end{align}
\end{subequations}
where $\eta^t = \varrho\cdot \eta^{t - 1}$ and $0<\varrho<1$.

\noindent $\blacklozenge$ For subproblem~\eqref{IPDD_p_1}, we have
\begin{equation} \label{IPDD_p_1_sub}
	\begin{split}
		\min_{ \bp } &~ \left\| \bp - \hat\bp \right\|_2^2 + \frac{1}{2\eta^t} \left\| \bp - \widetilde{\bz} \right\|_2^2    \\
		\mbox{s.t.}	
		&~ {\cal C}^{pz}: \ba_z^{\rm T}\bp \geq \gamma_{\rm th} \bb_z^{\rm T} \bp + \gamma_{\rm th} \sigma_k^2,
	\end{split}
\end{equation}
where $\widetilde{\bz} = \bz^t + \eta^t \bm\omega^t$, $\ba_z = \left[a_1z_1^t,a_2z_2^t,\dots,a_Kz_K^t\right]^{\rm T} \odot \be_k $, $\bb_z = \left[a_1z_1^t,a_2z_2^t,\dots,a_Kz_K^t\right]^{\rm T} \odot \left( \sum_{i=1,i\neq k}^{K}\be_i \right)$, and $\be_k = \left[ 0,\dots,1,\dots,0 \right]^{\rm T}$ is the indicator vector, where only the $k$-th element is 1 and others are 0.

If $\bp = \frac{2\hat\bp + \widetilde{\bz}/\eta^t}{2 + 1/\eta^t}$ satisfies constraint ${\cal C}_{pz}$, then the optimal solution of the problem~\eqref{IPDD_p_1_sub} is $\bp^\star = \frac{2\hat\bp + \widetilde{\bz}/\eta^t}{2 + 1/\eta^t}$.
Otherwise, the Lagrangian function of the problem~\eqref{IPDD_p_1_sub} is given by
\begin{equation*}
	\begin{split}
		{\cal L}_p = \left\| \bp - \hat\bp \right\|_2^2 + \frac{\left\| \bp - \widetilde{\bz} \right\|_2^2}{2\eta^t}  + \lambda_1 \left(\gamma_{\rm th} (\bb_z^{\rm T} \bp + \sigma_k^2) - \ba_z^{\rm T}\bp \right),
	\end{split}
\end{equation*}
where $\lambda_1\geq 0$ is a Lagrangian multiplier.

Therefore, we can derive the optimal solution as
\begin{equation} \label{optimal_p}
	\begin{split}
		\bp^\star &= \frac{2\hat\bp + \widetilde{\bz}/\eta^t + (\gamma_{\rm th} \bb_z - \ba_z)\lambda_1^\star }{2+1/\eta^t}.
	\end{split}
\end{equation}

Plugging~\eqref{optimal_p} back into the constraint ${\cal C}_{pz}$ such that the equality sign holds yields
\begin{equation} \label{optimal_lambda_1}
	\begin{split}
		\lambda_1^\star &= \frac{ \left( \gamma_{\rm th}\bb_z^{\rm T} - \ba_z^{\rm T} \right) \left( 2\hat\bp + \widetilde{\bz}/\eta^t \right) + \gamma_{\rm th} \sigma_k^2 (2+1/\eta^t)  }{\left( \gamma_{\rm th} \bb_z^{\rm T} - \ba_z^{\rm T} \right)  \left( \gamma_{\rm th} \bb_z - \ba_z \right) }.
	\end{split}
\end{equation}

\noindent $\blacklozenge$ For subproblem~\eqref{IPDD_p_2}, we have \vspace{-0.1cm}
\begin{equation} \label{IPDD_p_2_sub}
	\begin{split}
		\min_{ \bz } ~ \left\| \bz - \hat\bz \right\|_2^2 ~~
		\mbox{s.t.}	
		~ \ba_p^{\rm T}\bz \geq \gamma_{\rm th} \bb_p^{\rm T} \bz + \gamma_{\rm th} \sigma_k^2,
	\end{split}
\end{equation}
\vspace{-0.55cm}

\noindent where \vspace{-0.1cm}
$$\hat\bz = \bp^{t+1} - \eta^t \bm\omega^{t},\ba_p = \left[a_1^{t+1}p_1,a_2^{t+1}p_2,\dots,a_K^{t+1}p_K\right]^{\rm T} \odot \be_k ,$$ 
$$\bb_p = \left[a_1^{t+1}p_1,a_2^{t+1}p_2,\dots,a_K^{t+1}p_K\right]^{\rm T} \odot \left( \sum_{i=1,i\neq k}^{K}\be_i \right).$$

\vspace{-0.15cm}
If $\bz = \hat\bz$ satisfies constraint ${\cal C}_{pz}$, then the optimal solution to the problem~\eqref{IPDD_p_2_sub} is $\bz^\star = \hat\bz$.
Otherwise, by the Lagrangian method, we can derive the optimal solution as
\begin{equation} \label{optimal_z}
	\begin{split}
		\bz^\star &= \hat\bz - \frac{ \left( \gamma_{\rm th} \bb_p^{\rm T} - \ba_p^{\rm T} \right)\hat\bz \left( \gamma_{\rm th} \bb_p - \ba_p \right) + \gamma_{\rm th}\sigma_k^2 \left( \gamma_{\rm th} \bb_p - \ba_p \right) }{\left\| \gamma_{\rm th}\bb_p - \ba_p \right\|_2^2}.
	\end{split}
\end{equation}

\vspace{-0.55cm}
\subsection{Optimize $\bm\phi^l$}
Problem~\eqref{AO_SRM_4} with respect to $\bm\phi^l$ can be reformulated as
\begin{equation} \label{WSRM_Qos_Phi}
	\begin{split}
		\min_{ \bm\phi^l }~ \bm\phi^{l,{\rm H}} \bB \bm\phi^l - 2\Re\left\{ \bd^{\rm H} \bm\phi^l \right\} ~~ 
		\mbox{s.t.}~
		{\cal C}^\phi,~{\cal C}_k^R,~k\in {\cal K},
	\end{split}
\end{equation}
where $\bB = \sum_{i=1}^{K} \bC_{i}^{l,\rm H} \left[ \sum_{k=1}^{K} \zeta_k |u_k|^2 \bh_k \bh_k^{\rm H} \right] \bC_i^l$ and $\bd^{\rm H} = \sum_{k=1}^{K} \zeta_k u_k^* \bh_k^{\rm H} \bC_k^l  $.

By PDMM algorithm, the problem~\eqref{WSRM_Qos_Phi} is transformed into
\begin{equation} \label{WSRM_QoS_Phi_PDA}
	\begin{split}
		\min_{ \bm\phi^l } &~ \bm\phi^{l,{\rm H}} \bB \bm\phi^l - 2\Re\left\{ \bd^{\rm H} \bm\phi^l \right\} + \mu \left\| \bm\phi^l - \Pi_{{\cal C}^{\phi}} (\hat{\bm\phi}^l) \right\|_2^2  \\
		&~+ \mu \sum_{i=1}^{K} \left\|\bm\phi^l - \Pi_{{\cal C}_i^R}(\hat{\bm\phi}^l) \right\|_2^2.
	\end{split}
\end{equation}
Then, we can drive the optimal solution as
\begin{equation} \label{PDA_phi_solution}
	\begin{split}
		\bm\phi^l = \left(\bB + \mu(K+1)\bI_N \right)^{-1} ( \bd + \mu (\Pi_{{\cal C}^{\phi}}(\hat{\bm\phi}^l) + \sum_{i=1}^{K}\Pi_{{\cal C}_i^R}(\hat{\bm\phi}^l) ) ),
	\end{split}
\end{equation}
where for any $n\in {\cal N}$, we have
\begin{equation} \label{constant_module_proj}
	\left[\Pi_{{\cal C}^{\phi}}(\hat{\bm\phi}^l) \right]_n = 
	\begin{cases}
		\hat{\phi}_n^l , & \mbox{if } \big| \hat{\phi}_n^l \big| = 1,  \\
		\frac{\hat{\phi}_n^l}{| \hat{\phi}_n^l |}, & \mbox{otherwise}.
	\end{cases}
\end{equation}

Further, we propose efficient algorithms to derive the projection solution $\Pi_{{\cal C}_i^R}(\hat{\bm\phi}^l)$ for any $i\in {\cal K}$.
The projection solution $\Pi_{{\cal C}_k^R}(\hat{\bm\phi}^l)$ for any given $k$ can be obtained by solving the following problem
\begin{equation*}
	\begin{split}
		\min_{ \bm\phi } \left\| \bm\phi - \hat{\bm\phi}^l \right\|_2^2 ~
		\mbox{s.t.}	
		\frac{1}{\gamma_{\rm th}} \Re\left\{ \bm\phi^{{\rm H}} \bD_k \bm\phi \right\} \geq \Re\left\{ \bm\phi^{{\rm H}} \bE_k \bm\phi \right\} + \sigma_k^2,
	\end{split}
\end{equation*}
where 
$
\bD_k = \bC_{k}^{l,{\rm H}} \bh_k \bh_k^{\rm H} \bC_{k}^l;~
\bE_k = \sum_{i=1,i\neq k}^{K} \bC_{i}^{l,{\rm H}} \bh_k \bh_k^{\rm H} \bC_{i}^l.
$

Similarly, to solve the above problem, we introduce a new variable $\bm\kappa$ and transform it into
\begin{equation*} \label{WSRM_phi_proj_Ci_IPDD}
	\begin{split}
		\min_{ \bm\phi,\bm\kappa } &~ \left\| \bm\phi - \hat{\bm\phi}^l \right\|_2^2  \\
		\mbox{s.t.}	
		&~ \bm\kappa = \bm\phi,~ {\cal C}^{\phi\kappa}: \frac{1}{\gamma_{\rm th}} \Re\left\{ \bm\phi^{{\rm H}} \bD_k \bm\kappa \right\} \geq \Re\left\{ \bm\phi^{{\rm H}} \bE_k \bm\kappa \right\} + \sigma_k^2.
	\end{split}
\end{equation*}

Then, we use the IPDD algorithm to solve the above problem, which obeys the following updates \vspace{-0.15cm}
\begin{subequations}\label{IPDD_phi_Ci}
	\begin{align}
		\bm\phi^{t+1} = &~\mathop{\rm argmin}\limits_{\bm\phi \in {\cal C}^{\phi\kappa}} \left\| \bm\phi - \hat{\bm\phi}^l \right\|_2^2 + \frac{\left\| \bm\kappa^{t} - \bm\phi + \eta^t \bv^{t} \right\|_2^2}{2\eta^t} , \label{IPDD_phi_1} 
		\\
		\bm\kappa^{t+1} = &~\mathop{\rm argmin}\limits_{\bm\kappa \in {\cal C}^{\phi\kappa}} \left\| \bm\kappa - \bm\phi^{t+1} + \eta^t \bv^{t} \right\|_2^2, \label{IPDD_phi_2} \\
		\bv^{t+1} =&~\bv^{t} + (\bm\kappa^{t+1} - \bm\phi^{t+1}) \big/ \eta^t, \label{IPDD_phi_3}
	\end{align}
\end{subequations}
where $\eta^t = \varrho\cdot \eta^{t - 1}$ and $0<\varrho<1$.

For the subproblem~\eqref{IPDD_phi_1}, we have \vspace{-0.15cm}
\begin{equation} \label{IPDD_phi_1_proj}
	\begin{split}
		\min_{ \bm\phi } &~ \left\| \bm\phi - \hat{\bm\phi}^l \right\|_2^2 + \frac{1}{2\eta^t} \left\| \bm\phi - \tilde{\bm\kappa} \right\|_2^2   \\
		\mbox{s.t.}	
		&~\frac{1}{\gamma_{\rm th}} \Re\left\{ \bm\phi^{{\rm H}} \bD_k \bm\kappa^t \right\} \geq \Re\left\{ \bm\phi^{{\rm H}} \bE_k \bm\kappa^t \right\} + \sigma_k^2,
	\end{split}
\end{equation}
\vspace{-0.25cm}

\noindent where $\tilde{\bm\kappa} = \bm\kappa^t + \eta^t \bv^t$.

The Lagrangian function of the problem~\eqref{IPDD_phi_1_proj} is given by c
\begin{equation}
	\begin{split}
		{\cal L}_{\phi} = &\left\| \bm\phi - \hat{\bm\phi}^l \right\|_2^2 + \frac{1}{2\eta^t} \left\| \bm\phi - \tilde{\bm\kappa} \right\|_2^2 + \\
		&\lambda_2 \left(\Re\left\{ \bm\phi^{{\rm H}} \bE_k \bm\kappa^t - \frac{1}{\gamma_{\rm th}} \bm\phi^{{\rm H}} \bD_k \bm\kappa^t \right\} + \sigma_k^2 \right),
	\end{split}
\end{equation}
\vspace{-0.35cm}

\noindent where $\lambda_2\geq 0$ is a Lagrangian multiplier.

The corresponding Wirtinger derivative is given by \vspace{-0.1cm}
\begin{equation}
	\begin{split}
		\frac{\partial {\cal L}_{\phi}}{\partial \bm\phi^*} = \frac{(2\eta^t + 1) \bm\phi - \tilde{\bm\kappa}}{2\eta^t} - \hat{\bm\phi}^l + \lambda_2 ( \bE_k \bm\kappa^t - \frac{1}{\gamma_{\rm th}} \bD_k\bm\kappa^t ) = 0.
	\end{split}
\end{equation}

\vspace{-0.15cm}
Therefore, the optimal solution can be derived as \vspace{-0.1cm}
\begin{equation}\label{optimal_phi}
	\begin{split}
		\bm\phi = \frac{\hat{\bm\phi}^l + \tilde{\bm\kappa}/2\eta^t}{1 + 1/2\eta^t} - \frac{\bE_k \bm\kappa^t - \bD_k\bm\kappa^t/\gamma_{\rm th}}{1 + 1/2\eta^t}\lambda_2^\star.
	\end{split}
\end{equation}

\vspace{-0.15cm}
Plugging above expression to constraint such that the equality holds, the optimal $\lambda_2$ is given by \vspace{-0.1cm}
\begin{equation} \label{optimal_lambda_3}
	\begin{split}
		\lambda_2^\star = \frac{ \Re\{ (\hat{\bm\phi}^l + \frac{1}{2\eta^t}\tilde{\bm\kappa} )^{\rm H} \left(\gamma_{\rm th}\bE_k - \bD_k \right) \bm\kappa^t  \} + \sigma_k^2 \gamma_{\rm th}(1+\frac{1}{2\eta^t}) }{\Re\left\{(\bm\kappa^t)^{\rm H}\left( \bE_k - \frac{1}{\gamma_{\rm th}}\bD_k \right)^{\rm H} \left( \gamma_{\rm th} \bE_k - \bD_k \right) \bm\kappa^t  \right\}}.
	\end{split}
\end{equation}

\vspace{-0.15cm}
Similarly, the optimal solution of the problem~\eqref{IPDD_phi_2} can be derived as \vspace{-0.1cm}
\begin{equation} \label{optimal_kappa}
	\begin{split}
		\bm\kappa^\star &= \widetilde{\bm\phi} - \lambda_3^\star \left( \bE_k - \bD_k \right)^{\rm H} \bm\phi^{t+1}, \\
		\lambda_3^\star &= \frac{\Re\left\{(\bm\phi^{t+1})^{\rm H} \left( \gamma_{\rm th}\bE_k - \bD_k \right) \widetilde{\bm\phi} + \gamma_{\rm th} \sigma_k^2  \right\}  }{\Re\left\{(\bm\phi^{t+1})^{\rm H} \left(\gamma_{\rm th} \bE_k - \bD_k \right) \left( \bE_k - \bD_k/\gamma_{\rm th} \right)^{\rm H} \bm\phi^{t+1} \right\} },
	\end{split}
\end{equation}
\vspace{-0.25cm}

\noindent where $\widetilde{\bm\phi} = \bm\phi^{t+1} - \eta^t \bv^{t}$.

The overall algorithm for solving the SRM problem with QoS constraints is summarized in \textbf{Algorithm 2}.
\begin{algorithm}[t!]
	\caption{ Proposed Efficient VD-BSUM Algorithm  }
	\begin{algorithmic}[1]
		\STATE  {\bf Input:}  initialize $\bm\phi^{l,0}$, $\bp^0$, $\eta^0 = 100$, $\varrho=0.8$; $\ell=0$. 
		\REPEAT
		\STATE $u_k^{\ell+1} = \frac{ \bh_k^{\rm H} \bG^\ell \bw_k^1  p_k^\ell }{\sum_{i=1}^{K}| \bh_k^{\rm H} \bG^\ell \bw_i^1  |^2  (p_i^\ell)^2 + \sigma_k^2 },~k\in {\cal K}$; 
		\STATE $\zeta_k^{\ell+1} = \left( 1 - (u_k^{\ell+1})^* \bh_k^{\rm H} \bG^\ell \bw_k^1  p_k^\ell \right)^{-1},~k\in {\cal K}$; 
		\STATE {\bf{repeat}}
		\STATE $~$ $\left[ \Pi_{{\cal C}^{p} } (\hat\bp^{i}) \right]_k = {\rm max}\left( 0, \frac{\hat p^{i}}{1 + \rho^\star} \right)~,k\in {\cal K}$;
		\STATE $~$ {\bf{for $k$ from $1$ to $K$:}}
		\STATE $\quad$ {\bf{repeat}}
		\STATE $\qquad$ update $\tilde{\bp}^{t+1}$ according to~\eqref{optimal_p} and \eqref{optimal_lambda_1};
		\STATE $\qquad$ update $\bz^{t+1}$ according to~\eqref{optimal_z};
		\STATE $\qquad$ $\bm\omega^{t+1} = \bm\omega^{t} + (\bz^{t+1} - \tilde{\bp}^{t+1}) \big/ \eta^t$;
		\STATE $\qquad$ $ \eta^{t+1} = \varrho \cdot \eta^t$;
		\STATE $\qquad$ $ t = t + 1$;
		\STATE $~\quad$ {\bf{until}} $\left\| \tilde{\bp}^{t+1} - \bz^{t+1} \right\|_2^2 \leq 10^{-4}$;
		\STATE $~\quad$ $\Pi_{{\cal C}_{k}^R}(\hat\bp^i) = \tilde{\bp}^{t+1}$;
		\STATE $~$ {\bf{end}}
		\STATE $~$ update $\hat\bp^{i+1} $ according to~\eqref{closed_form_p};
		\STATE {\bf{until}} $\left\| \hat\bp^{i+1} - \frac{\sum_{k=1}^{K}\Pi_{{\cal C}_{k}^R}(\hat\bp^i) + \Pi_{{\cal C}^{p}}(\hat\bp^i)}{K+1} \right\|_2^2 \leq 10^{-4}$;
		\STATE $\bp^{\ell+1} = \hat\bp^{i+1}$;
		\STATE {\bf{for $l$ from $1$ to $L$:}}
		\STATE $~$ {\bf{repeat}}
		\STATE $\quad$ update $\Pi_{{\cal C}_{\phi}}(\hat{\bm\phi}^{l,i} )$ according to \eqref{constant_module_proj};
		\STATE $\quad$ {\bf{for $k$ from $1$ to $K$:}}
		\STATE $~\qquad$ {\bf{repeat}}
		\STATE $~\quad\qquad$ update $\bm\phi^{t+1}$ according to~\eqref{optimal_phi} and \eqref{optimal_lambda_3};
		\STATE $~\quad\qquad$ update $\bm\kappa^{t+1}$ according to~\eqref{optimal_kappa};
		\STATE $~\quad\qquad$ $\bv^{t+1} = \bv^{t} + (\bm\kappa^{t+1} - \bm\phi^{t+1}) \big/ \eta^t$;
		\STATE $~\quad\qquad$ $ \eta^{t+1} = \varrho \cdot \eta^t$;
		\STATE $~\quad\qquad$ $ t = t + 1$;
		\STATE $~\qquad$ {\bf{until}} $\left\| \bm\phi^{t+1} - \bm\kappa^{t+1} \right\|_2^2 \leq 10^{-4}$;
		\STATE $~\qquad$ $\Pi_{{\cal C}_{k}^R}(\hat{\bm\phi}^{l,i}) = \bm\phi^{t+1}$; 
		\STATE $\quad$ {\bf{end}}
		\STATE $\quad$ update $\hat{\bm\phi}^{l,i+1}$ according to~\eqref{PDA_phi_solution};
		\STATE $~$ {\bf{until}} $\left\| \hat{\bm\phi}^{l,i+1} - \frac{\sum_{k=1}^{K}\Pi_{{\cal C}_{k}^R}(\hat{\bm\phi}^{l,i}) + \Pi_{{\cal C}^{p}}(\hat{\bm\phi}^{l,i})}{K+1} \right\|_2^2 \leq 10^{-4}$;
		\STATE $~$ $\bm\phi^{l,\ell+1} = \hat{\bm\phi}^{l,i+1}$;
		\STATE {\bf{end}}
		\STATE ${\cal R}_{\rm sum}^\ell= \sum_{k}^{K} R_k $;
		\STATE $\ell= \ell+1$;
		\UNTIL  $\left| {\cal R}_{\rm sum}^{\ell+1} -  {\cal R}_{\rm sum}^{\ell} \right| \leq 10^{-4} $.
	\end{algorithmic}
\end{algorithm}

\vspace{-0.3cm}
\subsection{Computational Complexity Analysis}
In this subsection, we analyze the complexity of the proposed VD-BSUM algorithm in \textbf{Algorithm 2}.
The total complexity of the VD-BSUM algorithm is ${\cal O}( K^2MN + K^2T_p + L K N^2 T_\phi + LN^3 )$, where $T_p$ and $T_\phi$ denote the number of iterations when optimizing $\bp$ and $\bm\phi^l$, respectively.
This is because the overall complexity of the closed-form $\bu$, as well as $\bm\zeta$, is ${\cal O}(K^2 MN)$, the complexity for optimizing $\bp$ is ${\cal O}(K^2 T_p)$, and the complexity for optimizing $\{\bm\phi^l\}_{l=1}^L$ is ${\cal O}(LKN^2T_\phi + LN^3)$.

\begin{figure}[t]	
	\vspace{-0.4cm}
	\centering \includegraphics[width=0.8\linewidth]{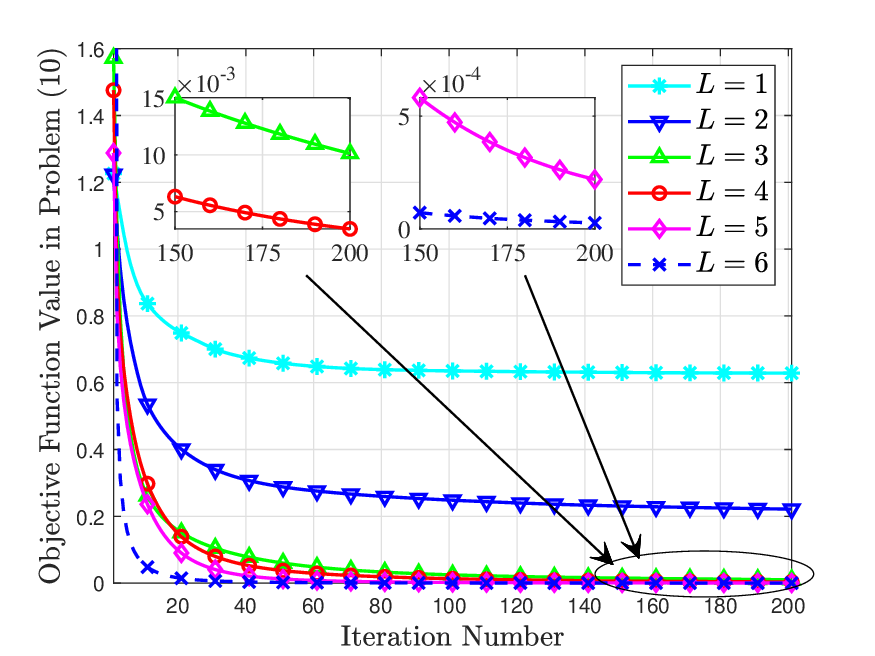}
	\vspace{-0.3cm}
	\caption{Convergence of the proposed AO-PDMM algorithm under different SIM layers.} 
	\vspace{-0.55cm}
	\label{fig:algorithm_convergence}
\end{figure}

{\it Remark 3:
	For problem~\eqref{WSRM_QoS}, the presence of QoS constraints significantly increases the challenges of finding an optimal solution.
	Common approaches reformulate the problem into a convex form using SCA or SDR algorithms to solve it.
	However, these methods typically incur high computational complexity.
	For these two benchmark algorithms, we first apply BSUM to reformulate the original problem~\eqref{WSRM_QoS} into the iterative form shown in~\eqref{AO_SRM}.
	1) For the benchmark SCA algorithm, a first-order Taylor approximation is employed to transform the constraint in~\eqref{AO_SRM_3} into a convex form, while the subproblem in~\eqref{AO_SRM_4} is still solved using the approach in~\eqref{WSRM_QoS_Phi_PDA}. The key difference is that, for each projection subproblem, all constraints are also approximated using first-order Taylor expansions to ensure convexity.
	2) For the benchmark SDR algorithm, we introduce a new variable $\widetilde\bPhi = \bm\phi^l \bm\phi^{l,\rm H}$ to reformulate the subproblem in~\eqref{AO_SRM_4} as a convex problem, and the rank-one solution is subsequently reconstructed using Gaussian randomization. 
	The subproblem in~\eqref{AO_SRM_3} is solved using the same approach.
}

\begin{figure*}[t]
	\centering
	\subfloat[desired beam pattern ]{\includegraphics[width=0.4\columnwidth]{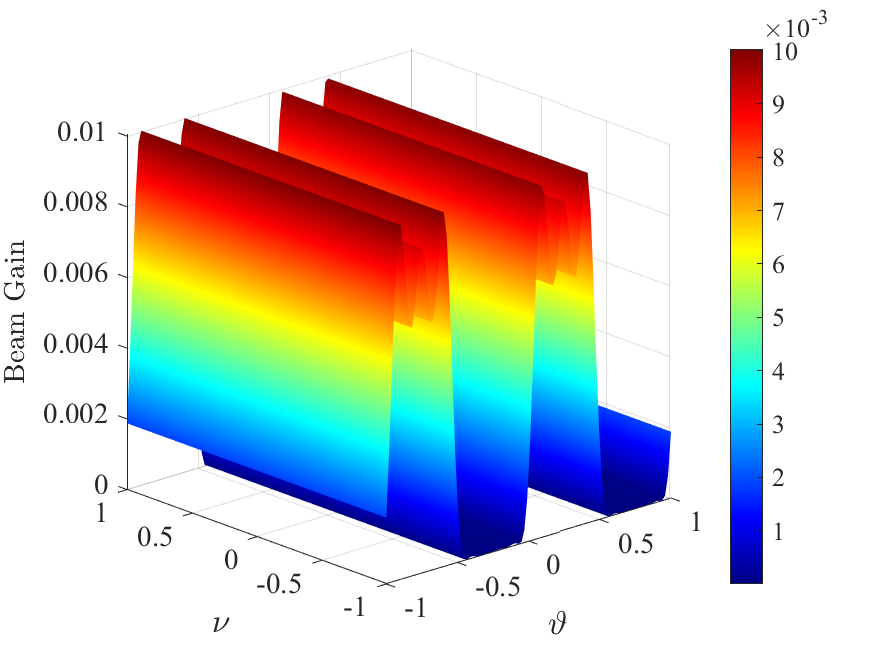}%
		\label{desired_beam_gain_theta1}}
	\hfil
	\subfloat[beam pattern ($L=1$) ]{\includegraphics[width=0.4\columnwidth]{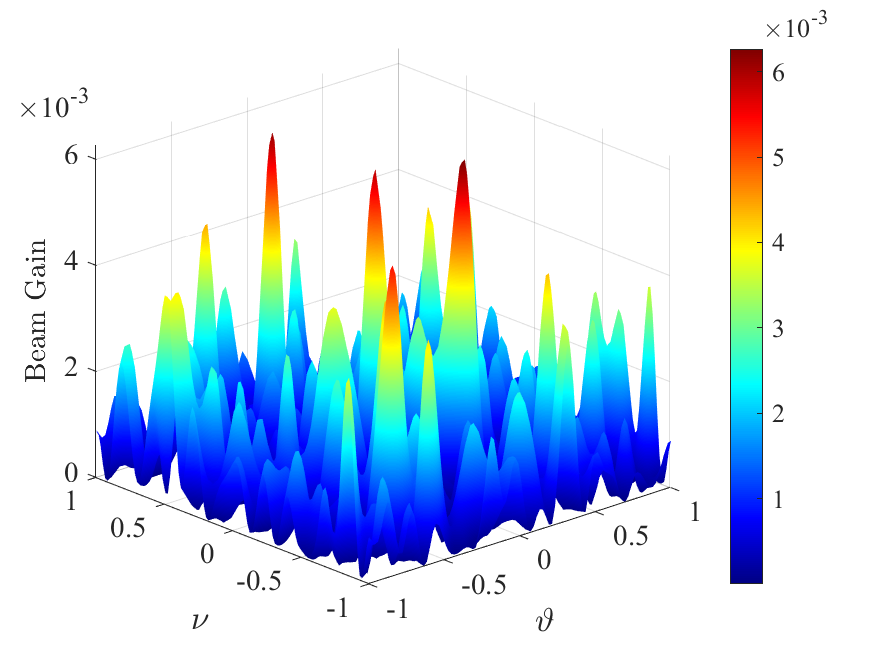}%
		\label{L1_beam_gain_theta1}}
	\hfil
	\subfloat[beam pattern ($L=2$) ]{\includegraphics[width=0.4\columnwidth]{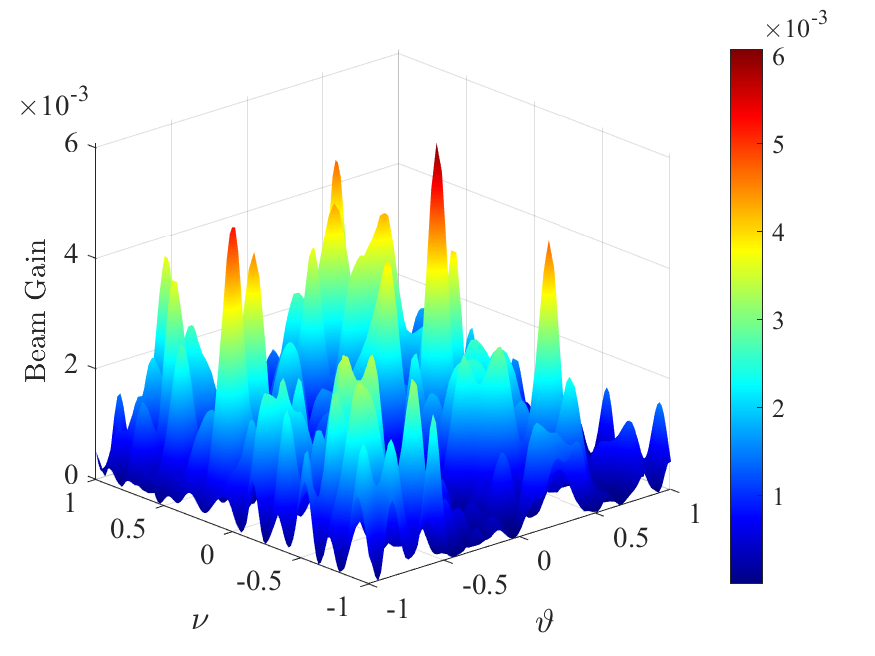}%
		\label{L2_beam_gain_theta1}}
	\hfil
	\subfloat[beam pattern ($L=4$) ]{\includegraphics[width=0.4\columnwidth]{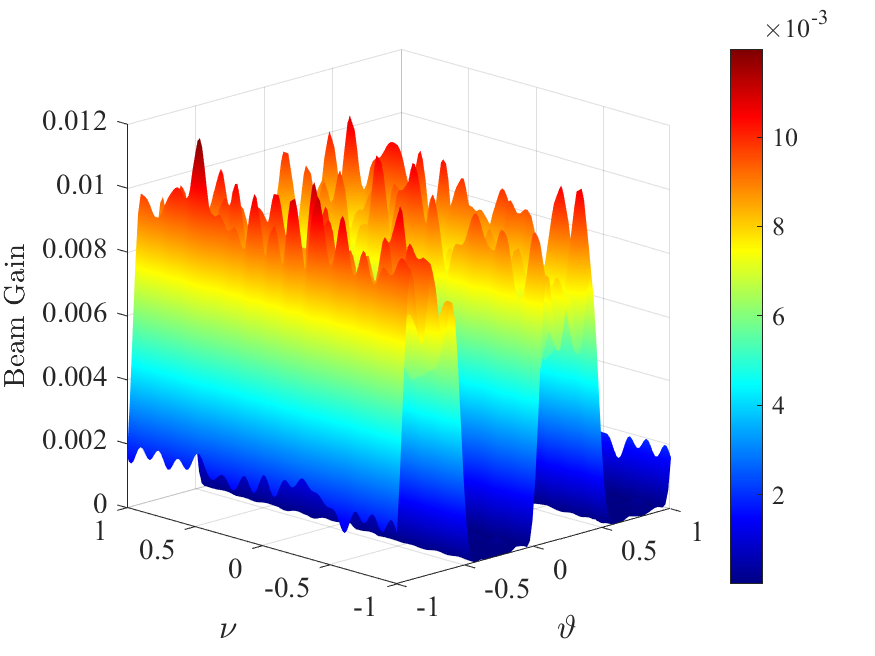}%
		\label{L4_beam_gain_theta1}}
	\hfil
	\subfloat[beam pattern ($L=7$) ]{\includegraphics[width=0.4\columnwidth]{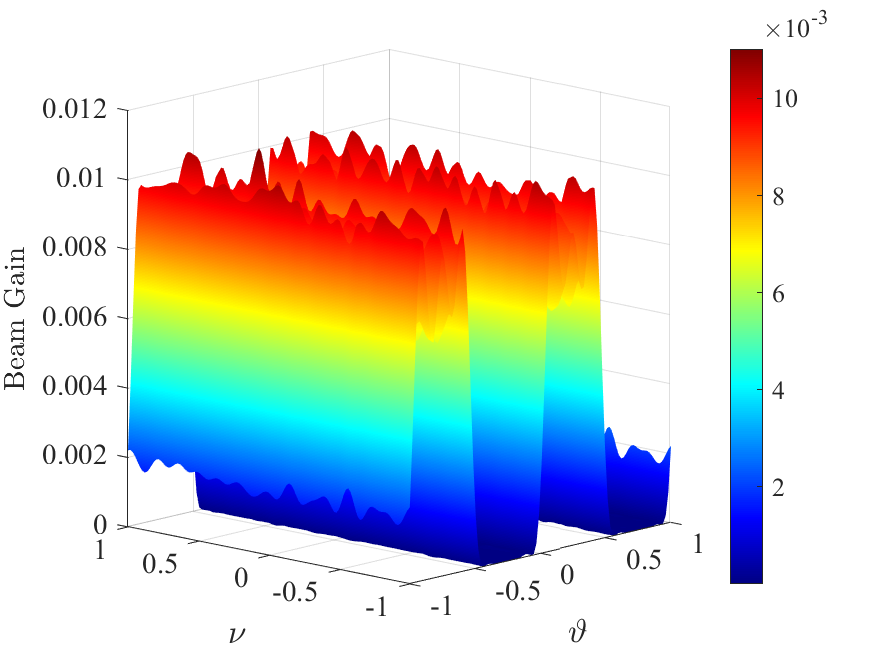}%
		\label{L7_beam_gain_theta1}}
	\hfil
	\centering
	\subfloat[desired beam pattern ]{\includegraphics[width=0.4\columnwidth]{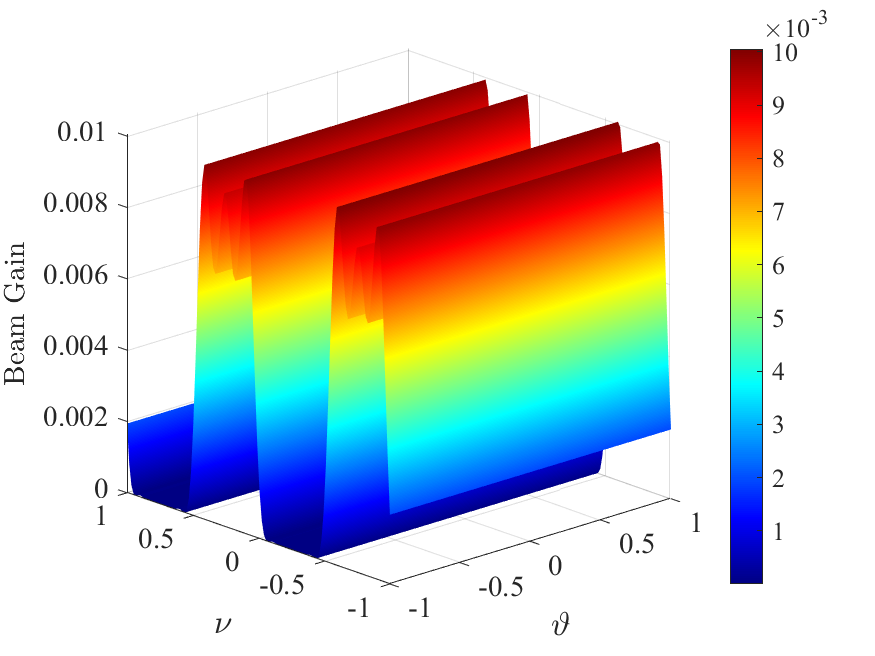}%
		\label{desired_beam_gain_theta2}}
	\hfil
	\subfloat[beam pattern ($L=1$)]{\includegraphics[width=0.4\columnwidth]{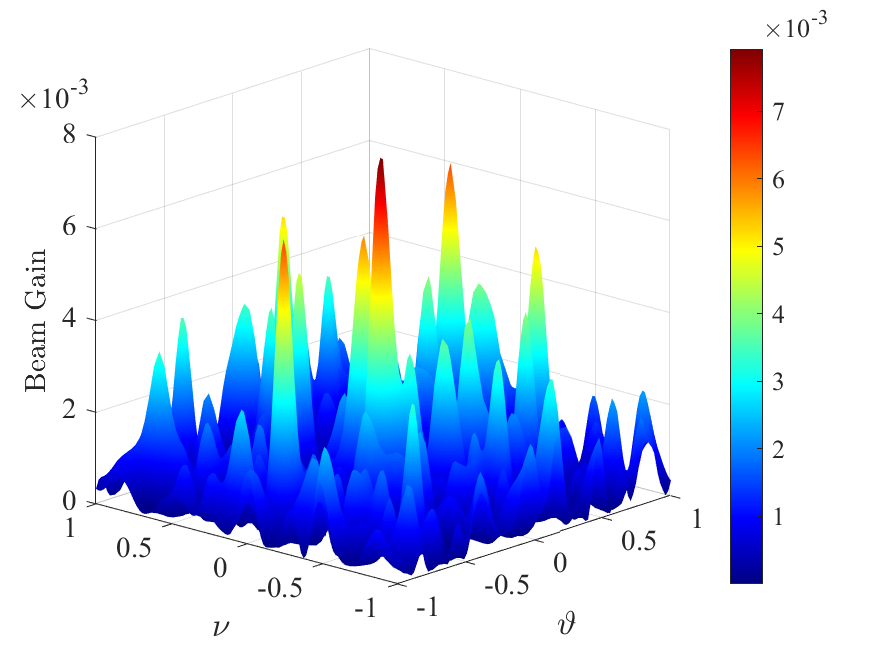}%
		\label{L1_beam_gain_theta2}}
	\hfil
	\subfloat[beam pattern ($L=2$)]{\includegraphics[width=0.4\columnwidth]{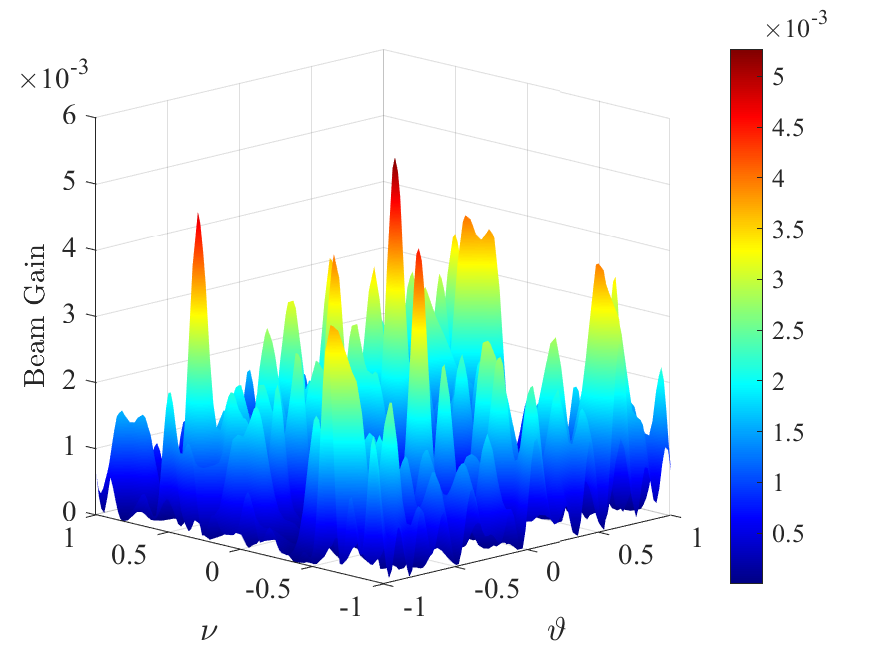}%
		\label{L2_beam_gain_theta2}}
	\hfil
	\subfloat[beam pattern ($L=4$)]{\includegraphics[width=0.4\columnwidth]{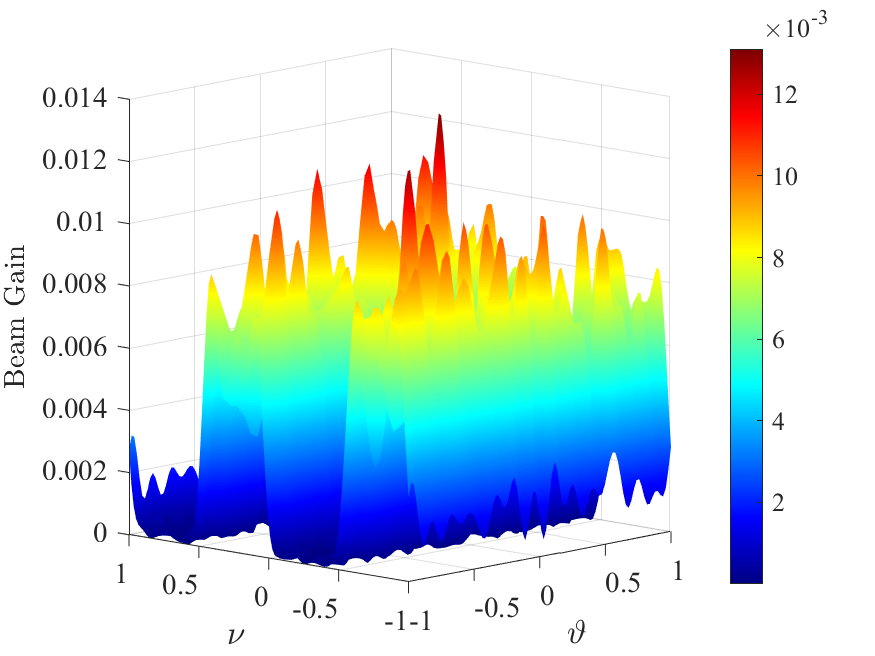}%
		\label{L4_beam_gain_theta2}}
	\hfil
	\subfloat[beam pattern ($L=7$)]{\includegraphics[width=0.4\columnwidth]{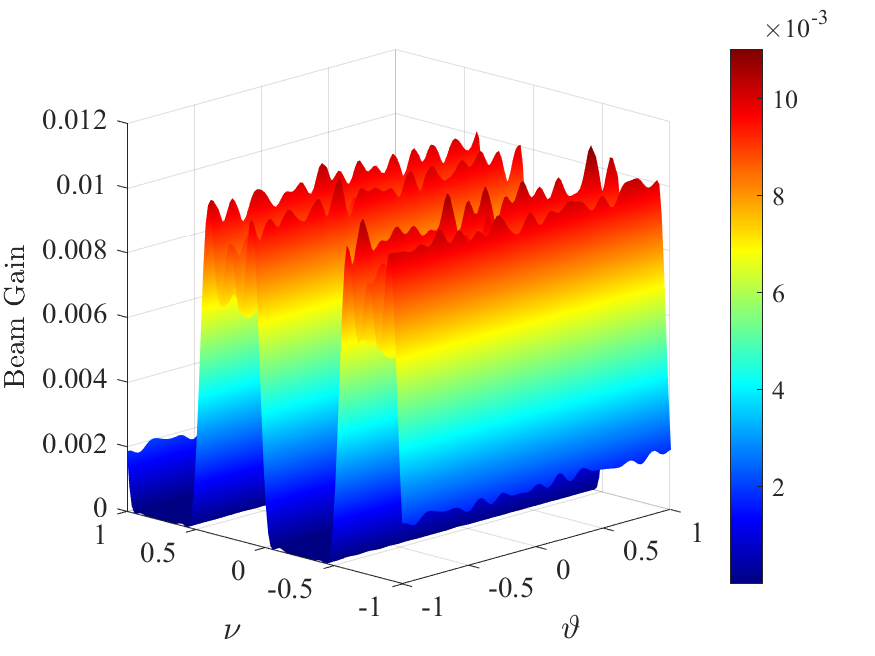}%
		\label{L7_beam_gain_theta2}}
	\vspace{-0.1cm}
	\caption{The beam pattern obtained by the proposed method under different SIM layers. $\vartheta$ direction: (a)(b)(c)(d)(e), $\nu$ direction: (f)(g)(h)(i)(j).}
	\vspace{-0.7cm}
	\label{fig:beam_pattern}
\end{figure*}
\begin{figure}[t]
	\centering
	\vspace{-0.25cm}
	\subfloat[Traditional EBT]{\includegraphics[width=0.45\columnwidth]{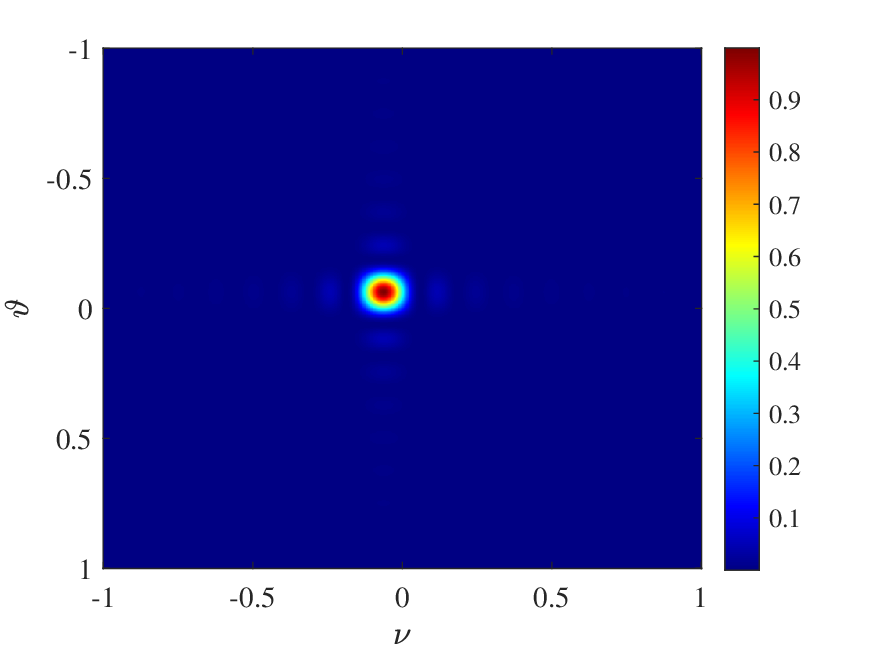}%
		\label{Exhuastive_beam_couple}}
	\hfil
	\subfloat[Proposed TLS-BT ]{\includegraphics[width=0.45\columnwidth]{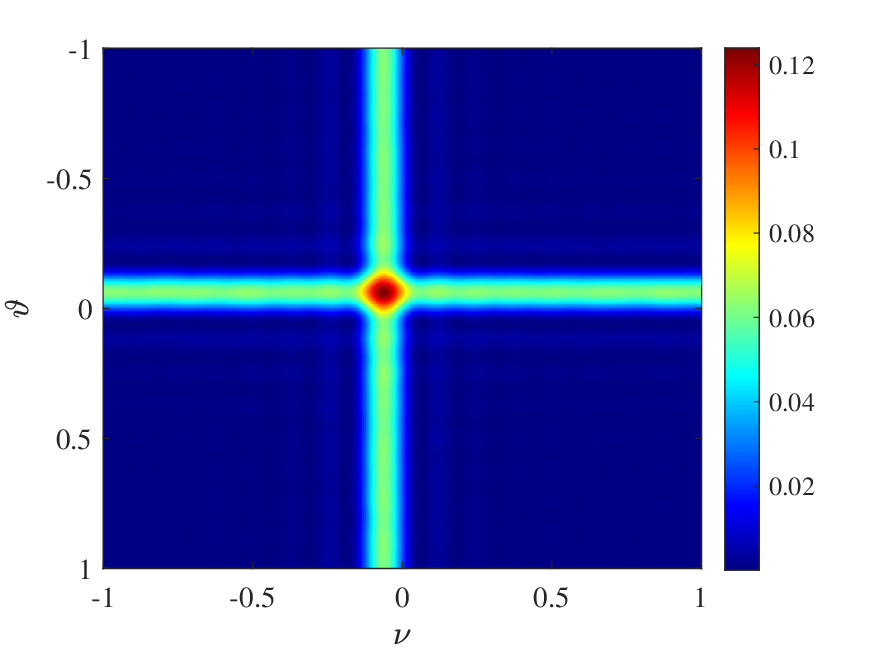}%
		\label{Exhuastive_beam_decouple}}
	\vspace{-0.1cm}
	\caption{Comparison of beam patterns between the traditional EBT and the proposed TLS-EBT, $L=7$.}
	\vspace{-0.55cm}
	\label{fig:Exhuastive_beam}
\end{figure}
\begin{figure}[t]	
	\centering \includegraphics[width=0.8\linewidth]{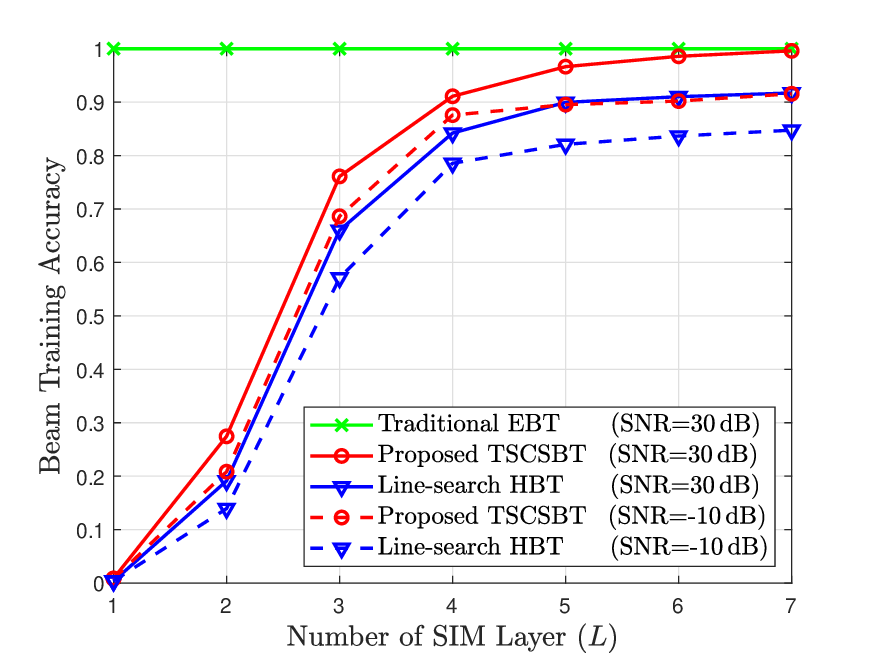}
	\vspace{-0.3cm}
	\caption{Beam training accuracy comparison of line-search HBT, traditional EBT, and proposed TSCSBT under different SIM layers.} 
	\vspace{-0.55cm}
	\label{fig:success_rate_L}
\end{figure}
\begin{figure}[t]	
	\centering \includegraphics[width=0.8\linewidth]{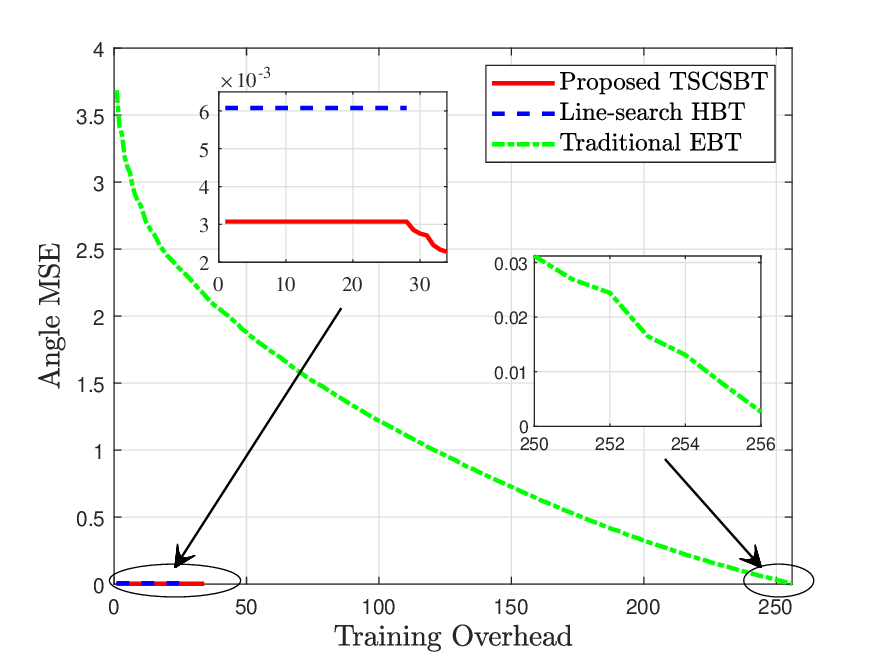}
	\vspace{-0.3cm}
	\caption{Comparison of the mean square errors (MSE) between the actual user angles and the estimated angles obtained by the proposed TSCSBT, the traditional HBT, and the traditional EBT methods, $L=7$, $\rm SNR=30\,$dB.} 
	\vspace{-0.5cm}
	\label{fig:MSE_Training}
\end{figure}
\begin{figure}[t]	
	\centering \includegraphics[width=0.8\linewidth]{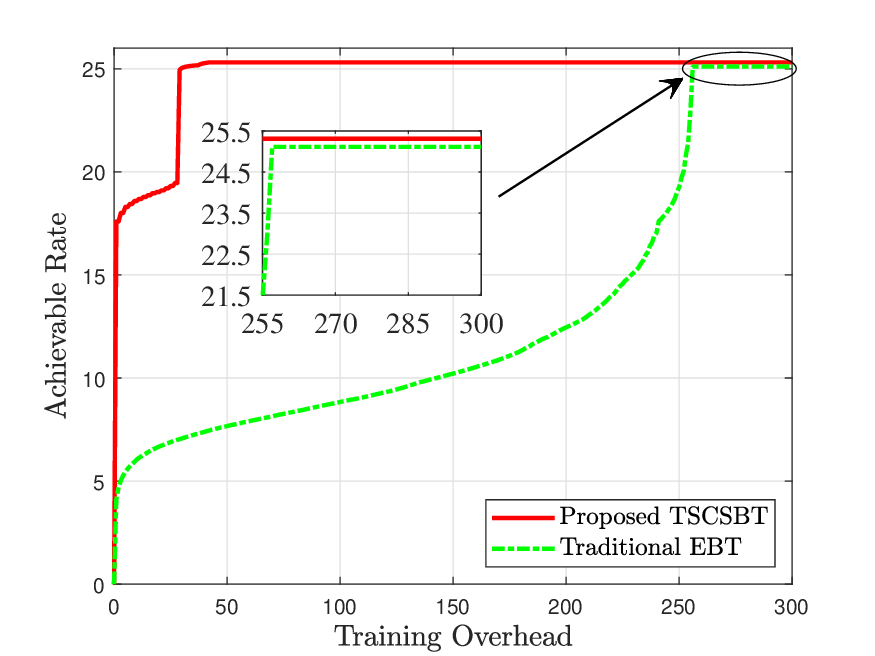}
	\vspace{-0.3cm}
	\caption{Comparison of the achievable rate achieved by the proposed TSCSBT and the traditional EBT as the overhead increases, $L=7$, $\rm SNR=30\,$dB.} 
	\vspace{-0.55cm}
	\label{fig:Rate_overhead}
\end{figure}

\vspace{-0.35cm}
\section{Simulation Results} \vspace{-0.1cm}
In this section, we present numerical results to demonstrate the effectiveness of the proposed beam training method and the superiority of the proposed VD-BSUM algorithm.
For simulations, we set the carrier frequency to 30GHz, the noise power $\sigma_k^2$ to -80dBm, the thickness of SIM to $5\lambda$, the large-scale fading coefficient to $\alpha_k=10^{-3}d_k^{-2.2}$, and $N_1=N_2=16$, where $d_k$ denotes the distance from SIM to user $k$.

\begin{figure}[t]
	\centering
	\vspace{-0.25cm}
	\subfloat[Angles of multi-path ]{\includegraphics[width=0.45\columnwidth]{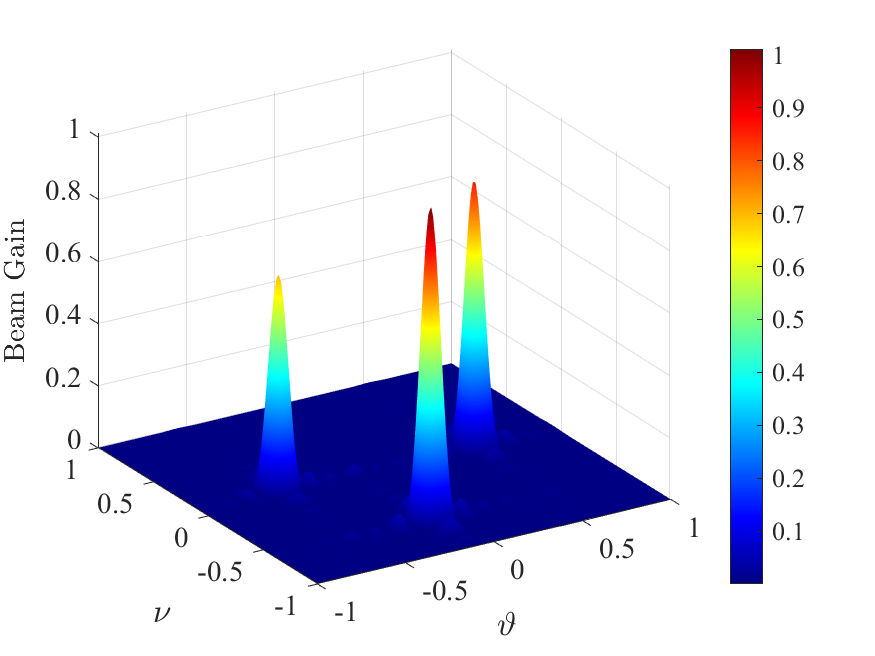}%
		\label{Multi_path_beam_pattern_practice}}
	\hfil
	\subfloat[Proposed MP-TSBT ]{\includegraphics[width=0.45\columnwidth]{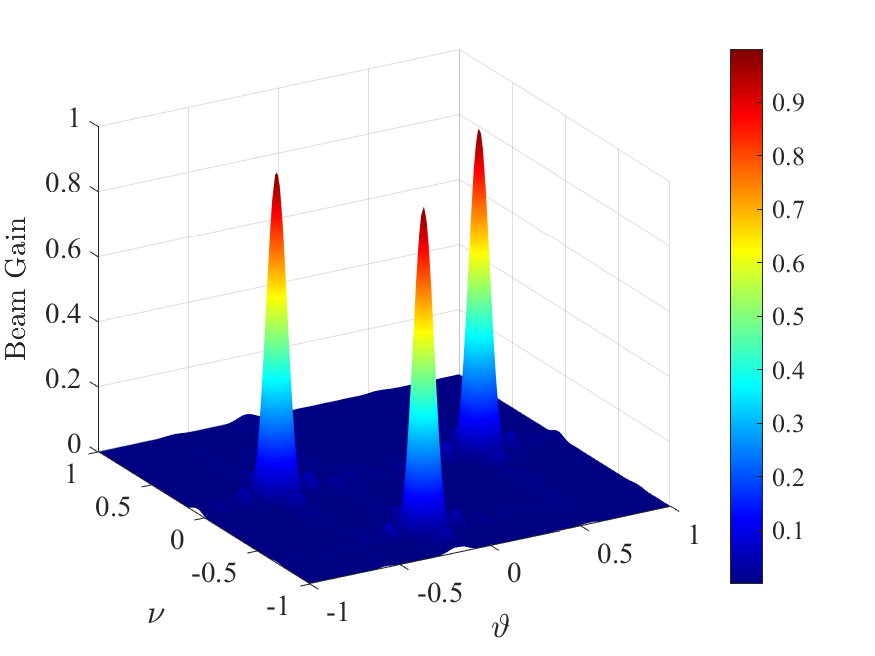}%
		\label{Multi_path_beam_pattern_training}}
	\vspace{-0.1cm}
	\caption{Comparison between the actual multipath angles of users and the angles estimated by the proposed MP-TSBT method, $L=7$.}
	\vspace{-0.5cm}
	\label{fig:Multi_path_beam_pattern}
\end{figure}
\begin{figure}[t]	
	\centering \includegraphics[width=0.8\linewidth]{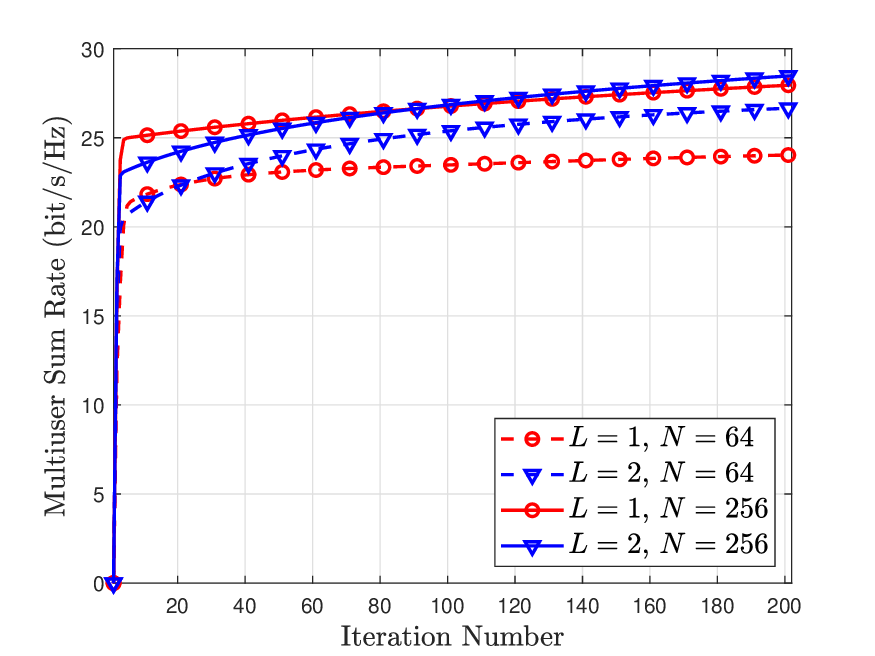}
	\vspace{-0.3cm}
	\caption{Convergence of the proposed VD-BSUM algorithm for solving the SRM problem with QoS constraints under different SIM layers, $P_{\rm max}=30\,$dBm, $K=3$. }
	\vspace{-0.4cm}
	\label{fig:algorithm_convergence_SRM}
\end{figure}
\begin{figure}[h!]
	\centering
	\subfloat[Average running time]{\includegraphics[width=0.8\columnwidth]{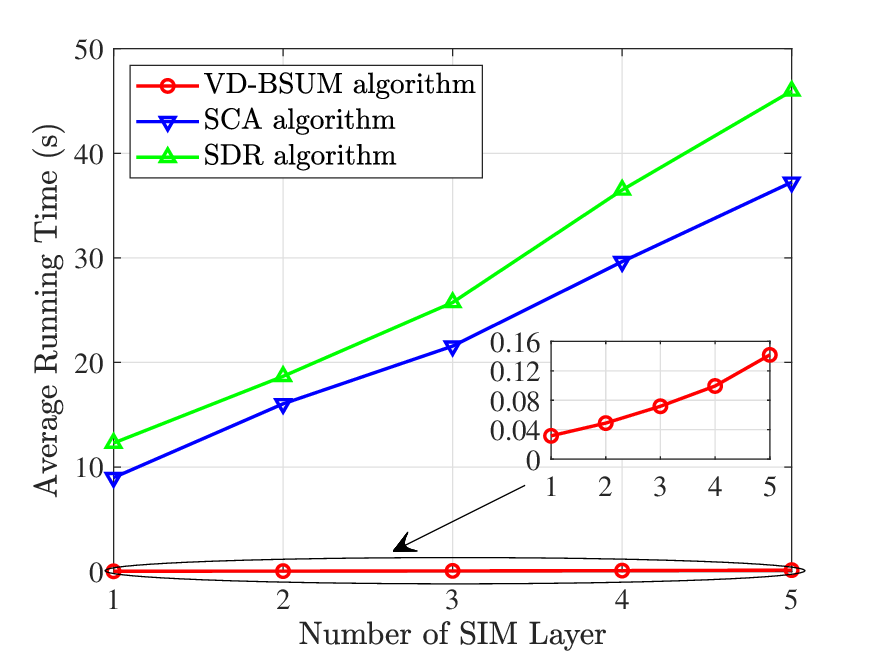}%
		\label{Algorithm_comparison_runtime}}
	\hfil 
	\vspace{-0.03cm}
	\subfloat[Multiuser sum rate ]{\includegraphics[width=0.8\columnwidth]{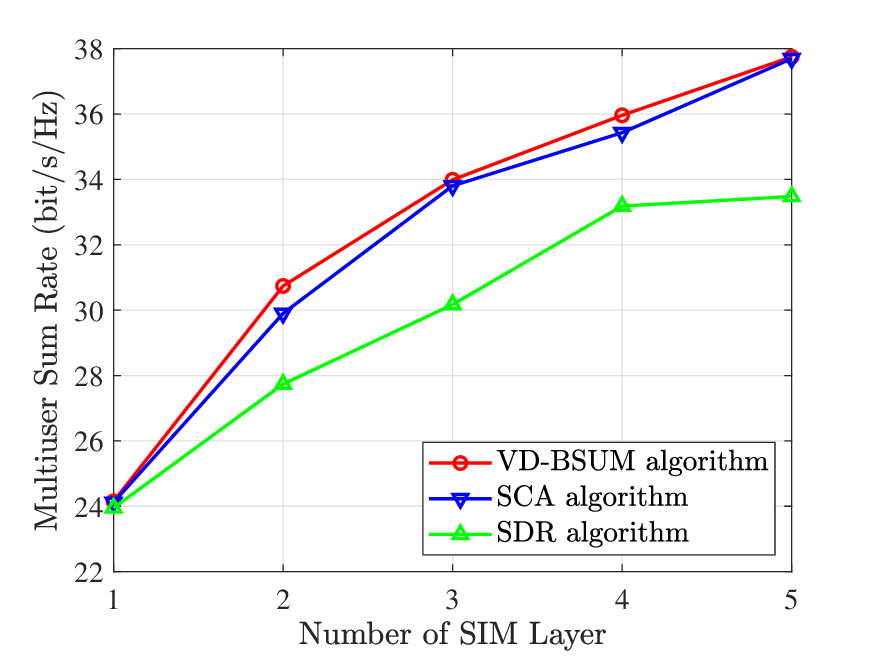}%
		\label{Algorithm_comparison_performance}}
	\vspace{-0.1cm}
	\caption{Comparison of proposed VD-BSUM algorithm, SCA algorithm, and SDR algorithm under different SIM layers, $P_{\rm max}=30\,$dBm, $K=3$, $N=8\times 8$.}
	\vspace{-0.6cm}
	\label{fig:algorithm_comparison}
\end{figure}
\begin{figure}[t]	
	\centering \includegraphics[width=0.8\linewidth]{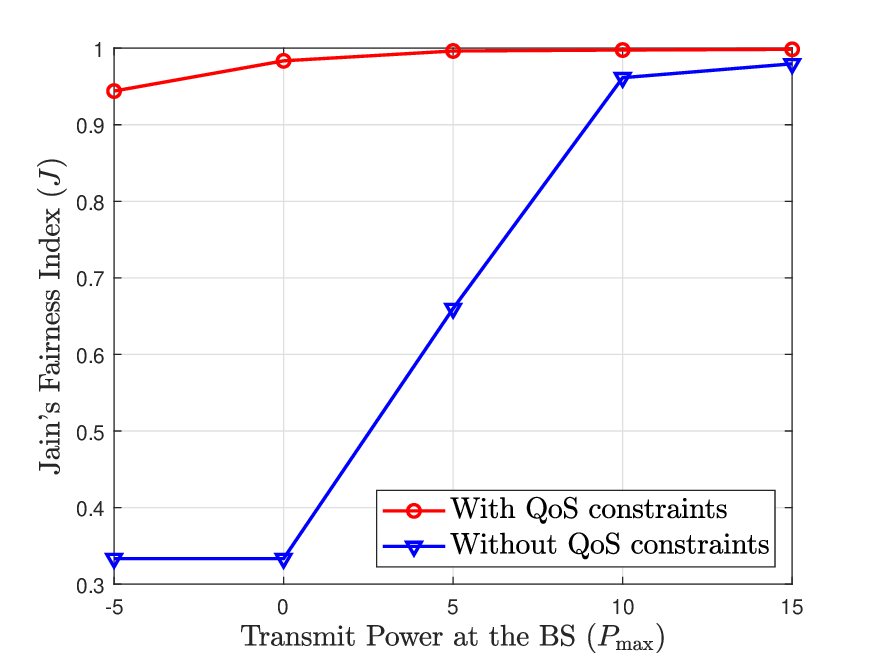}
	\vspace{-0.3cm}
	\caption{Comparison of user fairness obtained by solving the SRM problem with and without QoS for different transmitted powers respectively, $K=3$, $L=7$, $N=16\times 16$. }
	\vspace{-0.5cm}
	\label{fig:Jain_fairness_index}
\end{figure}
\begin{figure}[t]	
	\centering \includegraphics[width=0.8\linewidth]{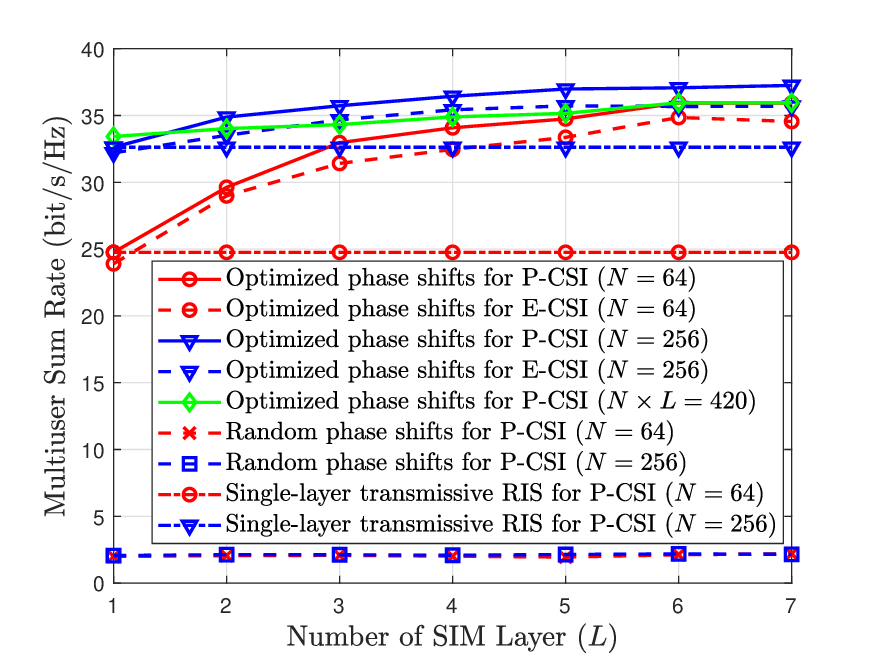}
	\vspace{-0.3cm}
	\caption{Multiuser sum rate comparison with different SIM layers and number of meta-atoms under P-CSI and E-CSI, $P_{\rm max}=30\,$dBm, $K=3$.} 
	\vspace{-0.5cm}
	\label{fig:Sum_rate_performance}
\end{figure}

\vspace{-0.4cm}
\subsection{Beam Training} \vspace{-0.1cm}
In this section, we evaluate beam training performance based on the constructed codebook.
First, we demonstrate the superior convergence behavior of the proposed AO-PDMM algorithm, as illustrated in Fig.~\ref{fig:algorithm_convergence}. 
It is observed that the objective function value decreases as the number of SIM layer $L$ increases, indicating that a SIM design with more layers enables a more accurate approximation of the desired beamforming vector.
Fig.~\ref{fig:beam_pattern}$\,$(a) and (f), obtained in the first step of the proposed TSCC method, demonstrate that the 2-D angle-domain decoupling approach enables accurate beam pattern design for individual angular dimensions.
Moreover, Fig.~\ref{fig:beam_pattern}$\,$(b)-(e) and (g)-(j) present the simulated beam patterns achieved by SIM with varying numbers of layers.
As $L$ increases, the realized beam patterns progressively converge to the desired beam pattern.
These results confirm that the proposed TSCC method can effectively construct SIM codebooks to support high-accuracy beam training in SIM systems.

We further compare the beam patterns obtained using the conventional 2-D grid search and the proposed 1-D linear search methods within the context of beam training.
Assuming the user is located at the angular direction $\vartheta = \nu = -\frac{1}{16}$, the goal is to form a high-gain beam precisely in that direction using the SIM codeword constructed through the proposed TSCC method when $L=7$.
As illustrated in Fig.~\ref{fig:Exhuastive_beam}$\,$(a),  the 2-D grid search method enables the generation of a well-focused directional beam at $\vartheta = \nu = -\frac{1}{16}$.
Fig.~\ref{fig:Exhuastive_beam}$\,$(b) demonstrates the beam patterns obtained using the codebook constructed by the proposed TSCC algorithm with angle-domain decoupling.
This approach allows for separate and independent detection along the $\vartheta = -\frac{1}{16}$ and $\nu = -\frac{1}{16}$ directions using a 1-D linear search along each angular dimension.
This effectively achieves the same directional sweeping capability as the 2-D grid search method but with significantly low complexity and training overhead.
These results highlight the practical advantages of the proposed TSCC-based beam training method in terms of both directional accuracy and implementation efficiency, making it well-suited for large-scale SIM systems.

Furthermore, to evaluate the effectiveness of the proposed TSCSBT method, we compare it with the HBT and conventional EBT schemes, where codebooks can be constructed through the TSCC method.
Since previous studies have not specified how to implement HBT in the 2-D angular domain, we adopt our proposed approach that decomposes the 2-D angular space into two independent 1-D angular domains. 
This allows HBT to be performed separately along each angular dimension, referred to as the {\it line-search HBT}, which serves as a benchmark for performance comparison.
As shown in Fig.~\ref{fig:success_rate_L}, we can see that the beam training accuracy gradually improves as the number of SIM layers increases, which is consistent with the conclusion in Fig.~\ref{fig:beam_pattern}.
The proposed TSCSBT method achieves significantly higher accuracy than line-search HBT, highlighting the robustness of TSCSBT to noise.
Additionally, the accuracy of CBT can be further improved by incorporating more advanced coding schemes with stronger error correction capabilities.
The beam training accuracy gradually increases as the SNR increases, and TSCSBT can still obtain superior accuracy gain compared to the line-search HBT.
When the SNR increases, the upper limit of the training accuracy that can be achieved with more SIM layers is higher, which can also be reflected in Fig.~\ref{fig:beam_pattern}.
Furthermore, we compare the angular estimation accuracy achieved by the three methods, using the mean square error (MSE) between the true and estimated angles as the evaluation metric. As shown in Fig.~\ref{fig:MSE_Training}, the proposed TSCSBT method attains higher angular estimation accuracy with lower training overhead compared to the conventional EBT, thanks to the introduced sliding beam training mechanism. In addition, TSCSBT also outperforms the line-search-based HBT in terms of angular accuracy, benefiting not only from the sliding beam training but also from its self-correction capability during the training process. Moreover, as illustrated in Fig.~\ref{fig:Rate_overhead}, the proposed TSCSBT achieves substantially higher rate performance than the conventional EBT while requiring significantly lower training overhead.
Fig.~\ref{fig:Multi_path_beam_pattern} illustrates the effectiveness of the proposed MP-TSBT scheme designed for multi-path detection.
In this scenario, we set $P=3$ and configure the threshold $P_{\rm th}$ to half of the peak power.
As observed, the proposed MP-TSBT accurately identifies and captures the user's multi-path components, demonstrating strong performance in multi-path environments.

\vspace{-0.3cm}
\subsection{SRM Problem with QoS Constraints}
From Fig.~\ref{fig:algorithm_convergence_SRM}, the superior convergence behavior of the proposed VD-BSUM algorithm in solving the SRM problem with QoS constraints is clearly demonstrated.
To further demonstrate the efficiency of the proposed algorithm, we compare the proposed  VD-BSUM algorithm, the SCA algorithm, and the SDR algorithms in terms of average running time and achieved multiuser sum rate performance.
As shown in Fig.~\ref{fig:algorithm_comparison}, our proposed VD-BSUM algorithm can achieve the lowest average running time while consistently delivering better performance than both the SCA and SDR algorithms.
Specifically, as illustrated in Fig.~\ref{fig:algorithm_comparison}$\,$(a), the computational efficiency advantage of the proposed VD-BSUM algorithm becomes more pronounced as the number of SIM layers $L$ increases compared to the SCA algorithm and the SDR algorithm.
More notably, even at $L = 1$, our proposed algorithm can still improve the computational efficiency gain by more than 2.8 times compared to the SCA and the SDR algorithms.

Furthermore, we emphasize the importance of incorporating QoS constraints, particularly when the available transmit power is limited. 
This is illustrated by comparing the user fairness index under the optimal solutions of the SRM problem with and without QoS constraints. 
User fairness is evaluated using Jain’s fairness index~\cite{Jain1984Fairness}, defined as
\[
J = \frac{\left( \sum_{k=1}^{K}R_k \right)^2 }{K \cdot \sum_{k=1}^{K} R_k^2 } \in \left[ \frac{1}{K}, 1 \right],
\]
where a larger $J$ indicates a fairer distribution of communication rates among users.
As shown in Fig.~\ref{fig:Sum_rate_performance}, when $P_{\rm max}$ is low, solving the SRM problem without QoS constraints results in unfair rate allocations, as the system favors a few strong users to maximize the overall sum rate. 
In contrast, incorporating QoS constraints significantly improves inter-user fairness, ensuring that system resources are more evenly distributed. 
This highlights the necessity of QoS-aware designs to prevent severe user discrimination in resource allocation, especially in power-limited scenarios.

In addition, to demonstrate the effectiveness of the proposed algorithm, we compare its performance under optimized phase shifts with that under randomly generated phase shifts. 
As shown in Fig.~\ref{fig:Sum_rate_performance}, the proposed algorithm ensures that the performance under optimized phase shifts is more than $12$ times better than that under random phase shifts.
Furthermore, beamformers designed using the user CSI estimated via the proposed beam training method (E-CSI) achieve a sum-rate performance that is very close to that obtained with perfect user CSI (P-CSI).
The SIM architecture also provides a notable improvement in sum-rate over conventional single-layer RIS designs.
Importantly, we demonstrate that, with a fixed total number of meta-atoms, increasing the number of layers (which reduces the number of elements per layer) can still lead to significant performance gains. In addition, the multiuser sum-rate rises with the number of SIM layers $L$ until it reaches saturation, indicating diminishing returns beyond a certain point. 
These results highlight the importance of proper system configuration and optimization to fully harness the benefits of SIM-based architectures.

\section{Conclusion}
In this paper, we presented a unified framework of low-complexity algorithms for SIM-assisted communication systems, addressing the challenges of channel state information acquisition and phase shift optimization. A generalized TSCC method was proposed, leveraging 2-D angular-domain decoupling for efficient SIM codebook design, with convergence guaranteed by the PDMM algorithm. To enable low-overhead and high-accuracy beam training, a TSCSBT method was developed and extended to multi-path user channels. Furthermore, a VD-BSUM algorithm was proposed to directly solve the QoS-constrained SRM problem via closed-form iterative updates with substantially reduced computational complexity. Simulation results validated that the proposed methods achieve accurate beam patterns, improved beam training accuracy and angular resolution, and enhanced sum-rate performance. The framework provides a practical and scalable solution for large-scale SIM-assisted systems and can be extended in future work to extremely large-scale near-field SIM communications.

\begin{appendices}
	\section*{Appendix A: Gap Analysis Between TSCC And Problem  \eqref{coded_codebook}}
	To quantify the optimality gap between the proposed TSCC method and Problem~\eqref{coded_codebook}, we proceed to derive an upper bound on the errors corresponding to the two problems.
	We first define $\bW_{\rm tscc}^\star = \bG^\star \bw^1$ and $\bW_{\rm orig}^\star = \bG_{\rm orig}^\star \bw^1$, where $\bG^\star$ and $\bG_{\rm orig}^\star$ are the optimal beamformer derived by the TSCC method and the problem~\eqref{coded_codebook}, respectively.

	Therefore, the residual of the problem~\eqref{coded_codebook} is given by
	\begin{equation*}
		\begin{split}
			E_{\rm orig} = \left\| \bA^{\rm H}\bW_{\rm orig}^\star - \bg_a\odot \bm\delta_a \right\|_2 \geq 0.
		\end{split}
	\end{equation*}
	Similarly, the residual of the TSCC method can be given by
	\begin{equation*}
		\begin{split}
			E_{\rm tscc} &= \| \underbrace{ \bA^{\rm H} \left(\bW_{\rm tscc}^\star - \bv_x^\star \otimes \bv_y^\star\right)}_{\text{step 2: measurement space error}}  + \underbrace{\bA^{\rm H} (\bv_x^\star \otimes \bv_y^\star) - \hat\bg}_{\text{step 1: propagated error}} \|_2 \\
			&\leq \left\| \bA \right\|_2 \epsilon_{\rm sim} + \left\|\bA_x^{\rm H} \bv_x^\star \right\|_2 \cdot E_y + \|\bg_y \|_2 \cdot E_x,
		\end{split}
	\end{equation*}
where $\hat\bg = \left( \bg_x \odot \bm\delta_x \right) \otimes \left( \bg_y \odot \bm\delta_y \right)$, $\epsilon_{\rm sim} = \|\bW_{\rm tscc}^\star - \bv_x^\star \otimes \bv_y^\star \|_2$, $E_y = \left\| \bA_y^{\rm H}\bv_y^\star - \bg_y \odot \bm\delta_y \right\|_2$, $E_x = \left\| \bA_x^{\rm H}\bv_x^\star - \bg_x \odot \bm\delta_x \right\|_2$.
	
	The desired beam gain is defined based on the desired beam pattern and can be set such that	$\bg_a\odot \bm\delta_a = \hat\bg$ and $\bA = \bA_x \diamond \bA_y$, where $\diamond$ denotes the Khatri-Rao production.
	Consequently, the residual $E_{\rm orig}$ is approximately equal to the residual in the first step of TSCC, and in general, the residual is a relatively small constant; that is,
	\begin{equation*}
		\begin{split}
	 		E_{\rm orig} \approx \left\|\bA_x^{\rm H} \bv_x^\star \right\|_2 \cdot E_y + \|\bg_y \|_2 \cdot E_x \leq \epsilon_{\rm constant},
		\end{split}
	\end{equation*}
	where $\epsilon_{\rm constant}$ is a given constant.

	Therefore, the gap between the proposed TSCC method and the problem~\eqref{coded_codebook} in codebook design can be characterized as follows
	\begin{equation}
		\begin{split}
			\Delta E = E_{\rm tscc} - E_{\rm orig} \leq \left\| \bA \right\|_2 \epsilon_{\rm sim}.
		\end{split}
	\end{equation}
	Based on this, it can be inferred that the performance gap between the proposed TSCC method and the problem~\eqref{coded_codebook} is determined by the representational capability of the SIM. Consequently, when the SIM has a sufficient number of layers and meta-atoms, the TSCC method can achieve a codebook design that closely approximates the solution of the problem~\eqref{coded_codebook}.

	The proof is complete. $\hfill\blacksquare$	
	
	\section*{Appendix B: Proof of PDMM Optimality}
	Consider the original problem
	\begin{equation} \label{original_problem}
		\begin{split}
			\min_{\bm\phi \in {\cal C}\triangleq \{ \bm\phi:|\phi_n|=1 \} }~\| \bC \bm\phi - \bv \|_2^2.
		\end{split}
	\end{equation}

	\vspace{-0.3cm}
	The associated KKT conditions of the problem~\eqref{original_problem} can be given by
	\begin{equation}\label{Orig_KKT}
		\begin{split}
			\bC^{\rm H}\left( \bC \bm\phi^\star - \bv \right) + \diag(\bm\lambda^\star)\bm\phi^\star = \bm 0, ~~
			|\phi_n^\star| = 1, 
		\end{split}
	\end{equation}
	where $\bm\lambda^\star \in \mathbb{R}^{N}$ is the dual variable. 
	
	The optimal solution obtained by the PDMM is given by
	\begin{equation} \label{PDMM_problem}
		\begin{split}
			\bm\phi^{(k+1)} = \arg\min_{\bm\phi}~\| \bC \bm\phi - \bv \|_2^2 + \mu^{(k)} \|\bm\phi - \bz^{(k)}  \|_2^2,
		\end{split}
	\end{equation}
	where $\bz^{(k)} = \Pi_{{\cal C}}(\bm\phi^{(k)})$. 
	
	Therefore, the closed-form solution can be derived as
	\begin{equation*}
		\begin{split}
			\bm\phi^{(k+1)} = \left( \bC^{\rm H} \bC + \mu^{(k)}\cdot \bI \right)^{-1} \left( \bC^{\rm H}\bv + \mu^{(k)} \bz^{(k)}  \right).
		\end{split}
	\end{equation*} 

\vspace{-0.3cm}
	Then, we can calculate the projection error as
	\begin{equation*}
		\begin{split}
			\|\bd^{(k+1)}\|_2 :=& \|\bm\phi^{(k+1)} - \bz^{(k)} \|_2
			\\
			=& \left\| \left( \bC^{\rm H} \bC + \mu^{(k)}\cdot \bI \right)^{-1} \bC^{\rm H}\left( \bv - \bC \bz^{(k)} \right) \right\|_2 \\
			\leq &  \frac{\|\bC \|_2 \left(\left\|\bv \right\|_2 + N \|\bC\|_2 \right)}{\mu^{(k)}} .
		\end{split}
	\end{equation*} 
	
	Therefore, by gradually increasing $\mu^{(k)}\rightarrow \infty$, the projection error can be guaranteed to convergence as follows
	\begin{equation} \label{projection_error}
		\begin{split}
			\|\bm\phi^{(k+1)} - \bz^{(k)} \|_2 \rightarrow 0.
		\end{split}
	\end{equation} 
	
	\vspace{-0.35cm}
	Moreover, since problem~\eqref{PDMM_problem} is convex and the projection error converges as in \eqref{projection_error}, both $\bm\phi^{(k)}$ and $\bz^{(k)}$ can converge to the optimal solution $\widetilde{\bm\phi}$ of the problem~\eqref{PDMM_problem}.
	Therefore, the optimal solution $\widetilde{\bm\phi}$ can satisfies the following conditions.
	\begin{equation}
		\begin{split}
			\bC^{\rm H}\left( \bC \widetilde{\bm\phi} - \bv \right) + \diag(\widetilde{\bm\lambda})\widetilde{\bm\phi} = \bm 0, ~~
			|\widetilde{\phi}_n| = 1, 
		\end{split}
	\end{equation}
	where $\widetilde{\lambda}_n = -\Re\left\{\widetilde{\phi}_n^* [ \bC^{\rm H} (\bC\widetilde{\bm\phi} - \bv) ]_n  \right\},~n=1,2,\dots,N$.
	
	The AO method~\eqref{AO_SIM} converges to a point satisfying the KKT conditions of the original problem~\eqref{SIM_codebook_problem}, since all blocks meet their first-order optimality at convergence.
	Hence, the proposed AO-PDMM method can converge to a locally optimal solution of the original problem~\eqref{AO_SIM}. 

	The proof is complete. $\hfill\blacksquare$

\end{appendices}

\bibliographystyle{IEEEtran}
\bibliography{refs}

\end{document}